\documentclass[12pt,nohyper]{JHEP3} 
\usepackage{epsfig}
\usepackage{psfrag} 
\usepackage{epstopdf}	
\def\freccia#1#2{{\stackrel{#2}{\hbox to #1{\rightarrowfill}}}}
\title{Type-IIA flux compactifications \\
and ${\cal N} = 4$ gauged supergravities} 
\author{ Gianguido Dall'Agata \\
Dipartimento di Fisica, Universit\`a di Padova and INFN, \\
Sezione di Padova, Via Marzolo 8, I-35131 Padova, Italy \\
E-mail: \email{gianguido.dallagata@pd.infn.it}} 
\author{ Giovanni Villadoro \\
Theory Group, Physics Department, CERN \\
CH--1211 Geneva 23, Switzerland \\
E-mail: \email{giovanni.villadoro@cern.ch}} 
\author{ Fabio Zwirner \\
Dipartimento di Fisica, Universit\`a di Padova and INFN, \\
Sezione di Padova, Via Marzolo 8, I-35131 Padova, Italy \\
E-mail: \email{fabio.zwirner@pd.infn.it}}
\preprint{CERN-PH-TH/2009-075 \\ 
DFPD-09/TH/10} 
\abstract{We establish the precise correspondence between Type-IIA flux compactifications preserving an exact or spontaneously broken ${\cal N} = 4$ supersymmetry in four dimensions, and gaugings of their effective ${\cal N} = 4$ supergravities.
We exhibit the explicit map between fluxes and Bianchi identities in the higher-dimensional theory and generalized structure constants and Jacobi identities in the reduced theory, also detailing the origin of gauge groups embedded at angles in the duality group.
We present AdS$_4$ solutions of the massive Type-IIA theory with spontaneous breaking to ${\cal N} = 1$, at small string coupling and large volume, and discuss their dual CFT$_3$.}
\keywords{Compactification and String Models, Flux Compactifications, D-branes, Supergravity Models}
\begin{document} 

\section{Introduction} \label{intro}

The relation between flux compactifications of higher-dimensional supergravities and gaugings of their effective four-dimensional theories has quite a long history \cite{ss}, with an extensive literature in the framework of superstring/M-theory compactifications (for a recent review and references to the original literature, see e.g.~\cite{reviews}).
When flux compactifications preserve an exact or spontaneously broken extended supersymmetry in four dimensions and there is a gap between the supersymmetry breaking scale and the compactification scale, the resulting gaugings are not only sufficient to fully determine the two-derivative low-energy effective Lagrangian, but also the only way in which a potential can be generated and some or all supersymmetries spontaneously broken.
While realistic four-dimensional effective theories have at most ${\cal N} = 1$ spontaneously broken supersymmetry~\footnote{Because of the chiral nature of weak interactions and of the direct and indirect evidence against  mirror fermions.}, in orientifold, orbifold and other string constructions a large amount of information can be extracted by the study of some underlying theory with $ {\cal N} > 1$.

In the present paper we concentrate on flux compactifications with exact or spontaneously broken ${\cal N} = 4$ local supersymmetry in four dimensions.
They are already quite well understood in the framework of heterotic \cite{nsw, fkpz, porzwi, kalmye} and Type-II compactifications \cite{FP,dfv,Berg:2003ri,aft, dkpz, dagfer,Aldazabal:2008zza}, but many open questions remain, especially in the framework of Type-IIA orientifolds, where the rich available structure of geometrical fluxes allows for interesting phenomena such as stable supersymmetric AdS$_4$ vacua (as found, for example, in some ${\cal N} = 1$ orbifolds \cite{dkpz, vzIIA, dWGKT,cfi}), and, perhaps, locally stable vacua with spontaneously broken ${\cal N} = 4$, $d=4$ supersymmetry and positive vacuum energy, even if no example was produced so far. 

The structure of our paper and its main results are described below.
In Section~2 we establish, in a quite general framework, the precise correspondence between Type-IIA flux compactifications preserving an exact or spontaneously broken ${\cal N} = 4$ supersymmetry and gaugings of their effective supergravities.
We focus on constructions with orientifold 6-planes (O6), in the presence of D6-branes parallel to the O6-planes and of general NSNS, RR and metric fluxes.
For simplicity, we neglect non-geometric fluxes and we consistently set to zero all brane-localized excitations, leaving these generalizations to future work.
We begin by recalling (following \cite{vzIIA, VZD}) some well-known properties of the chosen scheme for dimensional reduction: the field content of the effective theory, the allowed fluxes and the bulk and localized Bianchi Identities (BI).
We then recall the general structure of gauged ${\cal N} = 4$, $d = 4$ supergravity coupled to $n$ vector multiplets \cite{drw, schwei}, specializing to the case $n=6$ relevant for our discussion.
In particular, we recall the structure of the covariant derivatives acting on the scalar fields, the quadratic constraints on the gauging parameters, which play the role of generalized Jacobi identities, and the relation between the scalar potential and the supersymmetry variations of the fermionic fields.
We then spell out the precise correspondence between fluxes and BI of the compactified ten-dimensional theory on one side, generalized structure constants and Jacobi identities of the effective four-dimensional theory on the other side.
We confirm that, as implicitly introduced in \cite{implicit} and explicitly discussed in \cite{dkpz}, non-trivial duality phases (also known as de~Roo--Wagemans phases) \cite{drw} are generated.
We complete this section by discussing the role of a dilaton flux  to generate non-vanishing Sch\"on--Weidner parameters $\xi$ \cite{schwei} (in ${\cal N} = 4$ supergravity,  these parameters play a role analogous to Fayet--Iliopoulos terms in ${\cal N}=1$).

In Section~3 we apply our results and discuss the ${\cal N} = 4$ uplift of the family of ${\cal N} = 1$ AdS$_4$ supersymmetric vacua found in \cite{vzIIA}, performed by removing the $Z_2 \times Z_2$ orbifold projection used to reduce the amount of supersymmetry.
As a result, we find a family of Type-IIA AdS$_4$ vacua with spontaneous breaking of ${\cal N} = 4$ to ${\cal N} = 1$ and a number of adjustable free parameters.
These vacua \cite{Behrndt:2004km, Lust:2004ig} can be obtained without source terms, i.e.~with a vanishing net number of parallel D6-branes and O6-planes, guaranteeing that the ten-dimensional equations of motion are solved exactly. In the case of non-vanishing D6-brane
source terms the solution is still valid in the limit of smeared sources.
We comment on the associated geometry, on the uplift to ${\cal N} = 8$  obtained by removing the orientifold projection, and on the dual CFT$_3$ theories.
We conclude, in Section~4, with a brief discussion on possible generalizations and further applications of our results.
In the body of the paper, we make an effort to keep the technicalities to a minimum.
However, we find that some technical details on the symplectic embeddings may be useful to the supergravity specialists, thus we present them in the Appendix.


\section{Orientifold reduction and matching to ${\cal N}=4$} \label{general}

In this section we describe the reduction of Type-IIA supergravity on twisted tori orientifolds, where the orientifold involution acts non-trivially on three out of the six internal coordinates.
We allow for the presence of D6-branes parallel to the O6-planes, compatibly with ${\cal N} = 4$ supersymmetry, and for general NSNS and RR fluxes~\footnote{We do not consider non-geometric fluxes in this work, but we comment on some of the properties associated to turning on such deformations in section~\ref{sec:emvec}.}.
Since we are mainly concerned with the closed string sector, we only look at backgrounds with vanishing vacuum expectation values (vev) for the open string excitations, which would correspond to extra ${\cal N} = 4$ vector multiplets localized on the D6-branes.
The reduced theory is then a gauged ${\cal N} = 4$, $d = 4$ supergravity with six vector multiplets.
Our goal is to spell out the precise correspondence between fluxes and Bianchi Identities (BI) of the compactified ten-dimensional theory on one side, generalized structure constants and Jacobi identities of the effective four-dimensional theory on the other side. 

Here and in the following, we stick to the conventions of \cite{pol, vzIIA} unless otherwise stated. We will use $\mu$ and $i$ for the curved space-time indices corresponding to the four non-compact and the three compact dimensions parallel to the O6-planes world-volume, respectively, and $a$ for the three compact dimensions orthogonal to the O6-planes.

\subsection{Ten-dimensional fields, fluxes and constraints} 
\label{BI}

The bosonic NSNS sector of ${\it D} = 10$ Type-IIA supergravity consists of the (string-frame) metric $g$, the 2-form potential $B$ and the dilaton $\Phi$.
The intrinsic O6-parities are $+1$ for $g$ and $\Phi$, $-1$ for $B$.
After the O6 orientifold projection, the independent bosonic degrees of freedom in the NSNS sector of the reduced theory are the dilaton $\Phi$ and the following components of the metric and the $B$-field:
\begin{eqnarray}
	ds^2&=&g_{\mu\nu} \, dx^\mu dx^\nu+g_{ab} \, \eta^a \eta^b + g_{ij} (\eta^i+V_\mu^i dx^\mu) (\eta^j+V_\nu^j dx^\nu)\,,\nonumber \\
	B&=&B_{ai}\, \eta^a\, \eta^i\,, 
\end{eqnarray}
where here and in the following the wedge product is left implicit in antisymmetric forms.
The six internal 1-forms $(\eta^a,\eta^i$) satisfy the following relations: 
\begin{equation}
	\label{eq:defomegaeta} 
	\begin{array}{rcl}
	d\eta^k&=&\displaystyle \frac12 \, \omega_{ij}{}^k \, \eta^i \, \eta^j + \frac12 \, \omega_{ab}{}^k \, \eta^a\, \eta^b\,,\\[3	mm]
	d\eta^c&=&\omega_{ib}{}^c \, \eta^i\, \eta^b\,, 
	\end{array}
\end{equation}
which define the 9 ($\omega_{ij}{}^k$) + 9 ($\omega_{ab}{}^k$) + 27 ($\omega_{ib}{}^c$) metric fluxes.
The NSNS 3-form fluxes allowed by the O6 projection are (the numbers in brackets correspond to the multiplicities):
\begin{equation}
	\overline{H}_{abc} \;\; (1) \, , \quad \overline{H}_{ija} \;\; (9) \, . 
\end{equation}

The bosonic RR sector contains in principle the $p$-form potentials $C^{(p)}$ with $p=1,3,5,7,9$, whose intrinsic O6-parities are $+1$ for $p=3,7$ and $-1$ for $p=1,5,9$.
However, these degrees of freedom are not all independent, being related by Poincar\'e duality.
Before discussing how to identify the independent RR degrees of freedom that lead to the standard form of the effective ${\cal N} = 4$ supergravity, we display the field components that are invariant under the orientifold parity, organized in blocks of dual potentials, with their multiplicities in brackets:
\vspace{.1in}
\begin{equation}
	\begin{array}{r|ccccr|cc}
		{\scriptstyle{\rm scalars:}}\ &\ C^{(1)}_{a} \ & \ C^{(3)}_{i j k} \ & \ C^{(3)}_{i a b} \ & \ C^{(5)}_{i j a b c} \ & {\scriptstyle\rm vectors:\ } & \ C^{(5)}_{\mu i a b c} \ & \ C^{(3)}_{\mu a b} \ \\[3mm]
		&\updownarrow &\updownarrow &\updownarrow &\updownarrow &&\updownarrow &\updownarrow \\[3mm]
		{\scriptstyle \rm dual\ tensors:}\ &\ C^{(7)}_{\mu \nu ijk b c} \ & \ C^{(5)}_{\mu \nu a b c} \ & \ C^{(5)}_{\mu \nu j k c} \ & \ C^{(3)}_{\mu \nu k} \ &\qquad {\scriptstyle\rm dual\ vectors:\ } & \ C^{(3)}_{\nu j k} \ & \ C^{(5)}_{\nu i j k c} \ \\[5mm]
		&(3) & (1) & (9) & (3) & & (3) & (3) 
	\end{array}
	\label{rrpot} 
\end{equation}
\vspace{.1in}

\noindent
In summary, the bosonic RR sector contains 16 independent real degrees of freedom that can be described either by scalars or by 2-tensors, and 6 dual pairs of vectors.
Finally the candidate dual pairs of scalar and 4-tensor fluxes in the RR sector are
\vspace{.1in}
\begin{equation}
	\begin{array}{r|cccc}
	{\scriptstyle{\rm scalars:}}\	&	\ \overline{G}^{(0)} \ & \ \overline{G}^{(2)}_{i a} \ & \ \overline{G}^{(4)}_{i j a b} \ & \ \overline{G}^{(6)}_{i j k a b c} \ \\[3mm]
&\updownarrow &\updownarrow &\updownarrow &\updownarrow \\[3mm]
{\scriptstyle \rm dual\ tensors:}\ & \ \overline{G}^{(10)}_{\mu \nu \rho \sigma i j k a b c} \ & \ \overline{G}^{(8)}_{\mu \nu \rho \sigma j k b c} \ & \ \overline{G}^{(6)}_{\mu \nu \rho \sigma j k a b} \ & \ \overline{G}^{(4)}_{\mu \nu \rho \sigma} \ \\[5mm]
	& (1) & (9) & (9) & (1) 
	\end{array}
	\label{rrflu} 
\end{equation}

\vspace{.1in}

Our goal is, as in \cite{vzIIA}, to keep the scalar fields and to remove the 2-tensor fields, to keep the scalar fluxes and to remove the 4-tensor fluxes.
As we shall see, however, the presence of RR vectors in the $d = 4$, ${\cal N} = 4$ effective theory introduces additional complications: the vector combinations that must be kept will be identified later.

Summarizing, the bosonic field content of the reduced theory consists of 38 scalar degrees of freedom (22 from the NSNS sector, 16 from the RR sector) and 12 independent vector degrees of freedom (6 from the NSNS sector, 6 from the RR sector) in a suitable dual basis.

As it is well known, there are bulk and localized BI constraining the allowed systems of fields and fluxes. The first constraints come from the closure of the external derivative, $dd=0$, which, applied to eq.~(\ref{eq:defomegaeta}), implies the following constraints on the metric fluxes:
\begin{equation}
	\label{eq:omegaomega} \omega\,\omega= - \omega_{[mn}{}^q\,\omega_{p]q}{}^r =0 \,. 
\end{equation}
Notice that there are no localized source terms compatible with ${\cal N} = 4$ supersymmetry that can modify the above equations~\footnote{The KK5-monopoles discussed in \cite{VZKK} do preserve ${\cal N} = 4$ supersymmetry, but it is not the same ${\cal N} = 4$ supersymmetry preserved by the O6-planes.
Therefore, the AdS$_4$ vacuum discussed in \cite{dkpz} corresponds indeed to a gauged ${\cal N} = 2$ supergravity in the presence of the orientifold projection, and to a gauged ${\cal N} = 4$ supergravity only in the absence of the orientifold projection.}.
These however are not the only constraints that the metric fluxes must satisfy. The requirement that the compact six-manifold has no boundary corresponds to the constraint
\begin{equation}
	\label{eq:tromega} \omega_{mn}{}^n= 0 
\qquad
\Rightarrow
\qquad
\omega_{ik}{}^k+\omega_{ic}{}^c=0\,. 
\end{equation}
The general BI for $H$ in the absence of NS5-branes (which would break the ${\cal N} = 4$ supersymmetry) is simply
\begin{equation}
	d H = 0 \, , 
\end{equation}
whose solution can be written as
\begin{equation}
	H= d_4 B+\omega B + \overline H \, , 
\end{equation}
where we separated the various contributions: the derivative of the 2-form field $B$ with respect to the external coordinates (first term), the torsion term from the derivatives of the $\eta$ with respect to the internal coordinates (second term) and a constant flux term ($\overline H$), which must satisfy the integrability condition
\begin{equation}
	\label{eq:omegaH} \omega \ \overline H=0 \, . 
\end{equation}
In the absence of localized sources, the BI for the RR field strengths $G^{(p)}$ read
\begin{equation} \label{eq:RRBI}
	d G^{(p)} + H \, G^{(p-2)} = 0 \, , 
\end{equation}
and, in analogy with the previous discussion for $H$, the general solution for  $G^{(p)}$ is
\begin{equation}
	G^{(p)}= d_4 C^{(p-1)} + \omega \ C^{(p-1)} + H \, C^{(p-3)} + ( \mathbf{\overline G} \ e^{-B})^{(p)} \,, 
\end{equation}
where $\mathbf{\overline G}$ are constant fluxes subject to the integrability conditions
\begin{equation}
	\omega \ \overline G^{(p)} +\overline H \, \overline G^{(p-2)} =0 \,. 
\end{equation}
The last term in the solution is understood as expanded and projected into a $p$-form wedge product. The solution is valid in general, even when still keeping dual pairs of potentials, as long as there are no localized sources.
In the ${\cal N} = 4$ orientifold case under consideration, the only admissible localized sources are parallel D6-branes and O6-planes. The integrability condition for $G^{(2)}$ is then modified to
\begin{equation}
	\omega \ \overline G^{(2)} +\overline H \, \overline G^{(0)} = Q(\pi_6) \,,
\end{equation}
where $Q(\pi_6)$ is the sum of all Poincar\'e duals $[\pi_6]$ to the internal 3-cycles wrapped by the D6-branes and O6-planes. 
The presence of D6/O6 sources also implies further constraints that can be viewed as integrability conditions from the BI of localized fields.
In particular they read
\begin{equation}
	\overline H \, [ \pi_6 ] = 0 \, , \qquad \omega \ [ \pi_6 ] = 0 \,.  
\end{equation}
The first corresponds to the Freed--Witten \cite{FW} anomaly cancellation condition, which in our case is automatically satisfied, while the second (which is actually connected via dualities to the first) corresponds to requiring that the volume wrapped by the orientifold plane has no boundaries~\cite{VZD,Marchesano:2006ns,Villadoro:2007tb}. Explicitly the condition reads 
\begin{equation}
	\label{eq:constrtromega} \omega_{ik}{}^k=0\,, \qquad \omega_{ic}{}^c=0\,, 
\end{equation}
where the second equation follows from the first using eq.~(\ref{eq:tromega}).

\subsection{Effective ${\cal N}=4$ gauged supergravity}

The general structure of gauged ${\cal N} = 4$, $d = 4$ supergravity, with its gravitational multiplet coupled to $n$ vector multiplets, is known \cite{n4old, drw, schwei}. Its bosonic content consists of: the metric; $6 + n$ vector potentials $A_\mu^{M+}$ ($M=1,\ldots,6+n$), transforming in the fundamental vector representation of SO$(6,n)$ and carrying charge $+1$ under the SO$(1,1)$ subgroup of SU(1,1); the corresponding dual potentials $A_\mu^{M-}$, which also transform as a vector of SO$(6,n)$, but carry charge $-1$ under SO(1,1); $2+6 \, n$ real scalar fields, parameterizing the manifold
\begin{equation}
	\label{manifold} {\rm \frac{SU(1,1)}{U(1)} }\times \frac{{\rm SO}(6,n)}{{\rm SO(6) \times SO}(n)} \, . 
\end{equation}
Since we restrict ourselves to backgrounds with trivial open string vevs, from now on it will be sufficient to consider only the case $n=6$, neglecting the vector multiplets coming from D6-branes that act only as spectators.
According to \cite{schwei}, the complete Lagrangian is fully determined by two real constant tensors, $f_{\alpha \, MNP} = f_{\alpha \, [MNP]}$ and $\xi_{\alpha \, M}$, under the global on-shell symmetry group SU(1,1) $\times$ SO(6, 6), where $\alpha=+,-$ and $M=1,\ldots,12$. The index $M$ is lowered and raised with constant metric $\eta_{MN}$ and its inverse $\eta^{MN}$, whose explicit form will be given later.

The SU(1,1)/U(1) scalar manifold can be parameterized by the coset representatives
\begin{equation}
	{\cal V}_{\alpha}=\frac{1}{\sqrt{{\rm Im}\tau}}\left(
	\begin{array}{c}
		\tau \\
		1
	\end{array}
	\right) \, , \qquad (\alpha=+,-) \, , 
\end{equation}
where $\tau$ is a complex scalar field whose real and imaginary components are often called axion and dilaton, respectively.
In the gauged theory~\footnote{It is not restrictive to set all gauge coupling constants to one, by suitably rescaling the generalized structure constants $f$ and $\xi$.}, the covariant derivative of $\tau$ reads:
\begin{equation}
	\label{eq:defDtau} D_\mu \tau = 
	\partial_\mu \tau + A_\mu^{M-}\, \xi_{+M} + \left(A_\mu^{M+}\, \xi_{+M} -A_\mu^{M-}\, \xi_{-M}\right)\tau - A_\mu^{M+}\, \xi_{-M}\, \tau^2 \, . 
\end{equation}

The SO(6,6)/[SO(6) $\times$ SO(6)] scalar manifold can be parameterized by the coset representatives
\begin{equation}
	{\cal V} = \left( {\cal V}_M^{IJ} \, , \, {\cal V}_M^A \right) \, , 
\end{equation}
where $M=1,\ldots,12$ is a vector index of SO(6,6), $I, J = 1, \ldots, 4$ are indices in the fundamental representation of SU(4) $\sim$ SO(6) and $A=1,\ldots,6$ is a vector index of SO(6).
We exploit the fact that an SO(6) vector can alternatively be described by an antisymmetric tensor ${\cal V}^{IJ}={\cal V}^{[IJ]}$, subject to the pseudo-reality constraint
\begin{equation}
	{\cal V}_{IJ} = \left( {\cal V}^{IJ} \right)^* = \frac 1 2 \epsilon_{IJKL} {\cal V}^{KL} \; . 
\label{pseudoreality}
\end{equation}
The coset representatives must obey the constraint
\begin{equation}
	\eta_{MN} = - \frac 1 2 \epsilon_{IJKL} {\cal V}_M^{IJ} {\cal V}_N^{KL} + {\cal V}_M^A {\cal V}_N^A \; . 
	\label{cosetconstraint}
\end{equation}

The consistency of ${\cal N} = 4$ gaugings is enforced by a set of quadratic constraints on the generalized structure constants $\xi$ and $f$, which in turn can be interpreted as generalized Jacobi identities.
They read:
\begin{eqnarray}
	\xi_\alpha^M \xi_{\beta M} & = & 0 \, , \label{aa} \\[2mm]
	\xi^P_{(\alpha} f_{\beta)PMN} & = & 0 \, , \label{bb} \\[2mm]
	3 \, f_{\alpha R[MN} {f_{\beta PQ]}}^R + 2 \, \xi_{(\alpha[M} f_{\beta)NPQ]} & = & 0 \; , \label{cc} \\[2mm]
	\epsilon^{\alpha\beta} \left( \xi_{\alpha}^P f_{\beta PMN} + \xi_{\alpha M} \xi_{\beta N} \right) & = & 0 \, , \label{dd} \\[2mm]
	\epsilon^{\alpha\beta} \left( f_{\alpha MNR} {f_{\beta PQ}}^R - \xi^R_\alpha f_{\beta R[M[P} \eta_{Q]N]} - \xi_{\alpha[M} f_{N][PQ]\beta} + \xi_{\alpha[P} f_{Q][MN]\beta} \right) & = & 0 \, . \label{ee} 
\end{eqnarray}

A useful formula, against which we are going to fit the output of our generalized dimensional reduction, is the one giving the non-Abelian field strengths ${\cal H}^+$ in terms of the $A^+$ and $A^-$ potentials:
\begin{equation}
	{\cal H}_{\mu \nu}^{M \, +} = 2 \, 
	\partial_{[\mu} {A_{\nu]}}^{M \, +} - {\widehat{f}_{\alpha N P}}^{M} \, A_{[\mu}^{N \alpha} \, A_{\nu]}^{P \, +}  +\dots \,, \label{elfs} 
\end{equation}
where the dots refer to contributions from tensors, which cancel in the `electric' field strength combinations discussed later, 
and
\begin{equation}\label{DefThetaF} 
	\widehat{f}_{\alpha MNP} = f_{\alpha MNP} - \xi_{\alpha [M} \, \eta_{P] N} - \, \frac{3}{2} \, \xi_{\alpha N} \eta_{MP} \; . \end{equation}

To study the number of supersymmetries preserved by a given ground state, it is convenient to have explicit expressions for the supersymmetry variations of the fermions.
In the conventions of \cite{schwei}, the variations of the gravitino, dilatini and gaugini are given by
\begin{equation}
	\label{fermvar} \delta \psi_\mu^I = 2 D_\mu \epsilon^I - \frac{2}{3} \, A_1^{IJ} \, \Gamma_\mu \epsilon_J + \ldots \, , \quad \delta \chi^I = \frac{4}{3} \, i \, A_2^{IJ} \epsilon_J + \ldots \, , \quad \delta \lambda_A^I = 2 \, i \, {(A_{2A})_J}^I \, \epsilon^J + \ldots \; , 
\end{equation}
respectively, where~\footnote{We changed the convention for $A_{2\,A}{}^I{}_{J}$ and took the complex conjugate with respect to \cite{schwei}, to have all three $A$ matrices to act on the same SU(4) vector $q_I$.}
\begin{eqnarray}
	A_{1}^{IJ} & = & \epsilon^{\alpha\beta} \, {\cal V}_\alpha^\star \, {\cal V}^{M}_{KL} {\cal V}^{N \, IK} {\cal V}^{P \, JL} f_{\beta\,MNP} \label{eq:A1def} \, , \\[3mm]
	A_{2}^{IJ} & = & \epsilon^{\alpha\beta} {\cal V}_\alpha {\cal V}^{M}_{KL} {\cal V}^{N \, IK} {\cal V}^{P \, JL} f_{\beta\,MNP} + \frac 3 2 \, \epsilon^{\alpha \beta} \, {\cal V}_\alpha \, {\cal V}_M^{IJ} \, \xi_\beta^M \,, \label{eq:A2def} \\[3mm]
	{(\overline A_{2A})^I}_{J} & = & -\epsilon^{\alpha\beta} {\cal V}_\alpha^\star {\cal V}_A^M {\cal V}^{N \, IK} {\cal V}^P_{JK} \, f_{\beta\,MNP} - \frac 1 4 \, \epsilon^{\alpha \beta} \, {\cal V}_\alpha^\star \, {\cal V}_A^{M} \, \delta_J^I \, \xi_{\beta M} \, . \label{eq:A2AB} 
\end{eqnarray}
These expressions show that the $\xi_{\alpha M}$ act in a very similar way to Fayet--Iliopoulos parameters in ${\cal N}=1$ supergravity.
They do not appear in the mass matrix of the gravitini, eq.~(\ref{eq:A1def}), but provide a shift to the D-terms of eq.~(\ref{eq:A2def}).

Finally, the scalar potential $V$ is fixed in terms of the squares of the fermion variations by the following Ward identity of extended supergravity:
\begin{equation}
	\frac{1}{3} \, A_1^{IK } \, {\overline A}_{1\, JK} - \frac{1}{9} \, A_2^{IK} \, {\overline A}_{2\, JK} - \frac{1}{2} \, {A_{2\, AJ}}^K \, {\overline A}_{2\, A}{}^I{}_K \, = \, - \, \frac{1}{4} \, \delta^I_J \, V \, .\label{WardD4} 
\end{equation}

\subsection{Dimensional reduction from $d = 10$ to $d = 4$ with fluxes}

Since the $d = 4$ effective supergravity is completely determined, at the two-derivative level, by the gauging, we just need to focus on the effective action for the vector fields, from which we can read the couplings. First of all, we need to relate the zero modes of the ten-dimensional fields with the vectors $A^{M \pm}_\mu$. In our case the relations work as follows:
\begin{equation}	\label{eq:vectdef} 
	\begin{array}{llll}
A^{\overline{\imath} \, -}_{\mu } = {\widetilde V}_{\mu \, i}\,, &\displaystyle A^{i \, -}_{\mu} = \epsilon^{ijk} C^{(3)}_{\mu jk}\,,&\displaystyle A^{\overline a\, -}_{\mu } =\frac16 \epsilon^{ijk} C^{(5)}_{\mu a ijk}\,, &\displaystyle A^{a \, -}_{\mu} = \frac16\epsilon^{ijk} \epsilon^{abc} B^{(6)}_{\mu ijk bc}\,, 
		 \\&&&\\
A^{i \, +}_\mu = V_\mu^i\,, \qquad &\displaystyle A^{\overline{\imath}\, +}_{\mu} =\frac16 \epsilon^{abc} C^{(5)}_{\mu abc i}\,, \quad &\displaystyle A^{a \, + }_\mu =\frac12 \epsilon^{abc} C^{(3)}_{\mu bc}\,, \qquad & A^{\overline a \,+}_{\mu} = B_{\mu a}\,,
	\end{array}
\end{equation}
where the indices $M = (i , \overline{\imath} , a , \overline{a})$ in the fundamental vector representation of SO(6,6) are raised and lowered with the $12 \times 12$ constant metric
\begin{equation}
	\label{eq:defeta} \eta_{MN} = \eta^{MN} = {\bf{1}}_2 \otimes \sigma^1 \otimes {\bf{1}}_3 = \left( 
	\begin{array}{cccc}
		0 & {\bf{1}}_3 & 0 & 0 \\
		{\bf{1}}_3 & 0 & 0 & 0 \\
		0 & 0 & 0 & {\bf{1}}_3 \\
		0 & 0 & {\bf{1}}_3 & 0 
	\end{array}
	\right) \, . 
\end{equation}
Out of the 12+12 vector fields above, only 12 are independent.
In the ungauged case, we are completely free to choose the `electric' vectors, i.e.~the independent combinations of vectors that appear in the Lagrangian. When fluxes are turned on, however, the requirement of having an action written only in terms of scalar fields (without tensors) determines the electric and the magnetic combinations of vectors~\footnote{For a discussion of the role of tensor fields in gauged supergravities coming from flux compactifications and the relation between the standard and dual formulations see \cite{tens}.}.
If among the electric vectors entering the gauging both types of vector fields (those with positive and negative SO(1,1) charge) are present, the gauging is said to possess non-trivial duality phases, also known as de~Roo--Wagemans (dRW) phases.
The name `duality phases' follows from the fact that such a gauging corresponds to a non-trivial symplectic embedding of the gauge group inside the full duality group of symmetries of the ungauged theory, i.e.~an embedding providing an action of the gauge group where the vector field strengths and their duals get mixed (see \cite{dfv,aft,dagfer} for discussions of various ${\cal N} = 4$ cases coming from flux compactifications).
Since this is a technical point, we leave it for the Appendix.

In the following subsections we will first look at the covariant derivatives of the scalar fields, to find the `electric' combinations and identify the fluxes producing non-trivial dRW phases.
Then we will look at the covariant field strengths for the vectors, to read out the mapping between the fluxes and the structure constants of the gauging, which will fix the entire $d = 4$ action.

\subsubsection{Universal axion and SW parameters} \label{XI}

In our setup the universal axion (the one that, paired with a combination of the dilaton and of the O6 volume, reconstructs the complex scalar parameterizing the SU(1,1)/U(1) manifold) arises from the component of the RR 3-form potential parallel to the O6-plane, {\it viz.} 
\begin{equation}
	{\rm Re} \,\tau = \frac{1}{6} \epsilon^{ijk} C^{(3)}_{ijk} \, . 
\end{equation}
We can read off its covariant derivative by looking at the reduction of the corresponding RR 4-form on our background 
\begin{equation}
	D_\mu C^{(3)}_{ijk} = 
	\partial_\mu C^{(3)}_{ijk} - \omega_{[i l}{}^{l} \, C^{(3)}_{\mu j k ]} + V_\mu^h \omega_{h l}{}^{l} \, C^{(3)}_{ i j k } \, . 
\end{equation}
Comparing this expression with eq.~(\ref{eq:defDtau}), we see that the only components of $\xi_{\alpha M}$ that can be turned on in the chosen class of compactifications are $\xi_{+i}=\omega_{il}{}^{\, l}$. However, the constraint of eq.~(\ref{eq:constrtromega}) exactly forbids this possibility, thus it seems that no gaugings with non-trivial $\xi_{\alpha M}$ can be obtained from these string compactifications. In section~{\ref{dilflux}} we will comment on extensions that go around this limitation by
introducing a dilaton flux.

\subsubsection{Electric and magnetic vectors} \label{sec:emvec}

The `electric' vectors can be identified by looking at the combinations of vectors that appear in the covariant derivatives of the scalars. It is not difficult to see that the chosen set of fluxes does not produce gaugings involving the vectors dual to the metric and to the $B$-field, since in the NSNS sector all the scalars come from the dilaton, the metric and the $B$ field itself. In the RR sector, instead, scalars come from both $C^{(3)}$ and its dual $C^{(5)}$, therefore in general we expect that non-trivial combinations of the RR vectors and their duals can appear in the gauging.
We can thus restrict our analysis to the subset of 6+6 RR vectors and just look at the RR scalars.

As in the previous subsection, by looking at the reduction of the RR field strengths we can extract the relevant combinations:
\begin{eqnarray}
	\label{eq:Dscalars} D_\mu C^{(3)}_{abk} &=& 
	\partial_\mu C^{(3)}_{abk} +\omega_{ab}{}^l C^{(3)}_{\mu kl}+2\omega_{k[a}{}^d C^{(3)}_{\mu| b]d}+\dots \, ,   \\[3mm]
	D_\mu C^{(5)}_{abcij} &=& 
	\partial_\mu C^{(5)}_{abcij} +\omega_{ij}{}^k C^{(5)}_{\mu abc k}+\omega_{ab}{}^k C^{(5)}_{\mu c ijk} -{\overline H}_{abc} C^{(3)}_{\mu ij}-3{\overline H}_{ij[a} C^{(3)}_{\mu| bc]}+\dots \, , \nonumber
\end{eqnarray}
where the dots stand for contributions from NSNS vectors. Rewritten in terms of $d = 4$ supergravity vectors, these contributions can be conveniently summarized as
\begin{equation} \label{eq:summary}
	\begin{array}{c|cccc}
		& \  A^{i\,-}_\mu  \ & \  A^{a\,+}_\mu  \ & \  A^{{\overline i}\,+}_\mu  \ & \  A^{{\overline a}\,-}_\mu  \ \\[3mm]
		\hline \\[-2mm]
		\  C^{(3)}_{i a b}  \ & \  {\omega_{a b}}^{k}  \ &  {\omega_{i a}}^{b}  & 0 & 0 \\[3mm]
		\  C^{(5)}_{i j a b c}  \ & \  {\overline H}_{a b c}  &  {\overline H}_{i j c}  \ & \  {\omega_{i j}}^{k}  &  {\omega_{a b}}^{k}  \ 
	\end{array} \quad , 
\end{equation}
which shows the fluxes that determine what vectors (columns) enter the covariant derivative of each scalar (rows). The RR scalars are 12 (9 from $C^{(3)}$ and 3 from $C^{(5)}$), thus in principle we have 12 combinations of vectors in the covariant derivatives of the scalars. However, it can be shown that no more than six independent combinations of vectors are present.
To do this, it is enough to take the 12 magnetic combinations, obtained by dualizing those in eq.~(\ref{eq:Dscalars}), and to check that they are all orthogonal to the electric ones in eq.~(\ref{eq:Dscalars}). We have checked that this is indeed the case once the constraints of eqs.~(\ref{eq:omegaomega}), (\ref{eq:omegaH}) and (\ref{eq:constrtromega}) are imposed.

As it is obvious from eqs.~(\ref{eq:A1def}--\ref{WardD4}),  gaugings with non-trivial dRW phases are essential for moduli stabilization, since otherwise the SU(1,1)/U(1) scalar would enter homogeneously the scalar potential.
From (\ref{eq:summary}), we can see that the components $\omega_{a b}^{\quad k}$ and ${\overline H}_{a b c}$ are the only fluxes that involve vectors with negative SO(1,1) charge in the corresponding gauging. 
This is in agreement with \cite{dkpz}, which showed that exactly the same fluxes were responsible for producing a non-trivial dilaton dependence in the potential.

This result can be easily generalized to any ${\cal N} = 4$ orientifold compactification, including those with non-geometrical fluxes (${Q_m}^{qr}$, $R^{qrs}$) \cite{Shelton:2005cf}. 
Notice that all RR fluxes generate the same dRW phase, which can be set to zero by a suitable convention. Then, if we denote by $P^{qrs \ldots}_{mnp \ldots}$ the generic NSNS flux ($H_{mnp}$, ${\omega_{mn}}^q$, ${Q_m}^{qr}$, $R^{qrs}$), the rule-of-thumb reads: 
\vspace{10pt}

\emph{The NSNS fluxes leading to non-trivial dRW phases are those and only those with lower indices orthogonal to the O-planes and upper indices parallel to the O-planes}. \vspace{10pt}

For example, in the Type-IIB/O3 case, all $H$-fluxes give non-trivial dRW phases, since the indices are all orthogonal to the O3 planes, whereas all $Q$-fluxes give vanishing dRW phases. In the Type-IIA/O6 case, non-trivial dRW phases are generated by $H_{abc}$, ${\omega_{ab}}^i$, ${Q_a}^{ik}$, $R^{ijk}$.
In the Type-IIB/O9 case (and analogously in the heterotic case), all components of the $R$-fluxes (and only those) give non-trivial phases, since all internal indices are parallel to the O9-plane.

A similar reasoning applies to all the other cases, since by acting on an index with a T-duality in the corresponding direction, the dualized index is lowered or raised in the NSNS fluxes, but at the same time the corresponding direction changes from parallel to orthogonal to the O-planes, and viceversa.

In principle, for every flux we could also identify an S-dual flux \cite{Aldazabal:2006up}. 
Therefore, there should be other non-perturbative fluxes that generate non-trivial dRW phases.
In this case the rule just reverses, because by S-duality the SO(1,1) charge is inverted: S-dual NSNS fluxes always lead to non-trivial dRW phases except for those and only those with lower indices parallel to the O-planes and upper indices orthogonal to the O-planes. 
All S-dual RR fluxes give now non-vanishing dRW phases.  
This is in agreement with the results of \cite{Aldazabal:2008zza} for the Type-IIB/O3 case, where the authors show that structure constants with a negative SO(1,1) charge can be identified with non-trivial $H$-fluxes and with the S-dual of the non-geometric Q-fluxes.

\subsubsection{Gaugings from field-strength reduction}
\label{sub:Gaugings}

After having established that in the chosen compactifications it is always $\xi_{+M} = \xi_{-M} = 0$, our strategy to determine the remaining parameters of the ${\cal N} = 4$ gauging, i.e.~the generalized structure constants $f_{\alpha \, MNP}$, is to perform the dimensional reduction of the various field strengths in the NSNS and RR sectors, and to compare them with eq.~(\ref{elfs}). 

From the ten-dimensional Einstein term, adapting the results of \cite{ss} to our conventions, we obtain:
\begin{equation}
	V_{\mu\nu}^i=2 \, 
	\partial_{[\mu}V^{i}_{\nu]}-\omega_{ij}^{\ \ k} \, V^i_{\mu}\,V^j_{\nu} \, . 
\end{equation}
By reducing the NSNS 3-form field strength, the relevant terms read
\begin{equation}
	H_{\mu\nu a}=2 \, 
	\partial_{[\mu}B_{\nu]a}+2V^i_{[\nu} \omega_{ia}^{\ \ c}B_{c|\mu]} + V_{\mu}^iV_\nu^j {\overline H}_{ija} + \dots \, ,
\end{equation}
where, as before, the dots refer to contributions from tensor fields that cancel out when the `electric' vector-field 
combinations are considered.
In the RR sector, we have to consider the 4-form and 6-form field strengths, namely
\begin{eqnarray}
	G^{(4)}_{\mu\nu ab}&=&2
	\partial_{[\mu} C^{(3)}_{\nu]ab}-2{\overline G}^{(0)} B_{[\mu|a}B_{\nu]b} +2V^i_{[\nu} \left[\omega_{ab}^{\ \ k}C^{(3)}_{k|\mu]i}+\omega_{ia}^{\ \ c}C^{(3)}_{c|\mu]b} +\omega_{bi}^{\ \ c}C^{(3)}_{c|\mu]a}\right. \nonumber \\
	&& \left. +2B_{\mu][a}G^{(2)}_{i|b]}\right]+V^i_\mu V^j_\nu {\overline G}^{(4)}_{ijab}+\dots\,, 
\end{eqnarray}
\begin{eqnarray}
	G^{(6)}_{\mu\nu i abc}&=&2
	\partial_{[\mu} C^{(5)}_{\nu]iabc}+2\left(\omega_{ia}^{\ \ d} B_{d [\mu}C^{(3)}_{\nu]bc}+2\ {\rm Permut}_{abc}\right) \nonumber\\[2mm]
	&& -2\left( {\overline G}^{(2)}_{ia} B_{[\mu|b}B_{\nu]c}+2\ {\rm Permut}_{abc} \right) \nonumber \\[2mm]
	&& -2 V^j_{[\nu}\left[\omega_{ij}^{\ \ k}C^{(5)}_{k|\mu]abc} +\left(\omega_{ab}^{\ \ k}C^{(5)}_{\mu]cijk}+2\ {\rm Permut}_{abc}\right)  \right.  \\[2mm]
	&& \left.-{\overline H}_{abc} C^{(3)}_{\mu]ij} -\left( {\overline H}_{ija} C^{(3)}_{\mu]bc}+2\ {\rm Permut}_{abc} \right) - \left({\overline G}^{(4)}_{ijab}B_{\mu]c} +2\ {\rm Permut}_{abc}\right)\right] \nonumber \\[2mm]
	&& +V^j_\mu V^k_\nu {\overline G}^{(6)}_{ijkabc}+\dots\, . \nonumber
\end{eqnarray}
where the symbol ``$2\ {\rm Permut}_{abc}$'' stands for the two combinations obtained by cyclic permutation of the indices $abc$ of
the preceeding term.
Identifying the vector fields with the combinations having a definite SO(1,1) charge,  given previously in eqs.~(\ref{eq:vectdef}), we obtain:
\begin{eqnarray}
	V_{\mu\nu}^i&=&2 \, 
	\partial_{[\mu}A^{+\,i}_{\nu]}-\omega_{ij}^{\ \ k} \, A^{+\,i}_{\mu} A^{+\, j}_{\nu} \, , \\[3mm]
	H_{\mu\nu a}&=&2 \, 
	\partial_{[\mu}A^+_{\nu]a}+2\omega_{ia}^{\ \ c}A^+_{[\mu|c}A^{+i}_{\nu]}+ {\overline H}_{ija} A_{\mu}^{+\,i} A_\nu^{+\,j} + \dots \, , \\[2mm]
	\frac12\epsilon^{abc}G^{(4)}_{\mu\nu ab}&=&2
	\partial_{[\mu} A^{+c}_{\nu]}-{\overline G}^{(0)} \epsilon^{abc} A^+_{\mu a} A^+_{\nu b} + \frac12\omega_{ab}^{\ \ k} \epsilon^{abc}\epsilon_{ijk} A^{-\,i}_{\mu} A^{+\,j}_{\nu} +2\omega_{ia}^{\ \ c} A^{+\,a}_{[\mu} A^{+\,i}_{\nu]} \nonumber \\[2mm]
	&& -2{\overline G}^{(2)}_{ia}\epsilon^{abc} A^{+}_{[\mu|\,b} A^{+\,i}_{\nu]} + \frac12{\overline G}^{(4)}_{ijab} \epsilon^{abc} A^{+\,i}_{\mu} A^{+\,j}_{\nu}+\dots\,, \\[3mm]
	\frac16 \epsilon^{abc} G^{(6)}_{\mu\nu abci}&=&2
	\partial_{[\mu} A^{+}_{\nu]i} +2 \omega_{ia}^{\ \ c} A^+_{[\mu|c} A^{+\,a}_{\nu]} +\epsilon^{abc} {\overline G}^{(2)}_{ia} A^+_{\mu\,b}A^+_{\nu\,c} +2\omega_{ij}^{\ \ k} A^+_{[\mu| k} A^{+\,j}_{\nu]} \nonumber \\[2mm]
	&& -\frac12\omega_{ab}^{\ \ k} \epsilon^{abc} \epsilon_{ijk} A^-_{[\mu| c} A^{+\,j}_{\nu]} +\frac16{\overline H}_{abc}\epsilon^{abc}\epsilon_{ijk} A^{-\,j}_{[\mu} A^{+\,k}_{\nu]} -2{\overline H}_{ija} A^{+\,a}_{[\mu} A^{+\,j}_{\nu]}\nonumber \\[2mm]
	&& -{\overline G}^{(4)}_{ijab}\epsilon^{abc} A^+_{[\mu|c}A^{+\,j}_{\nu]} -\frac16 {\overline G}^{(6)}_{ijkabc}\epsilon^{abc}A^{+\,j}_{\mu} A^{+\,k}_{\nu}+\dots \, . 
\end{eqnarray}
We can now read the relation between fluxes and generalized structure constants by comparing with eq.~(\ref{elfs}):
\begin{eqnarray} \label{strcon} 
	f_{-\,ijk}&=&-\frac16 {\overline H}_{abc}\,\epsilon^{abc}\, \epsilon_{ijk}\,, \nonumber \\[2mm]
	f_{-\,ij}^{\ \ \ \ c}&=&- \frac12 {\omega}_{ab}^{\ \ k}\,\epsilon^{abc} \,\epsilon_{ijk}\,, \nonumber \\[2mm]
	f_{+}^{\ abc}&=& {\overline G}^{(0)}\, \epsilon^{abc} \,, \nonumber \\[2mm]
	f_{+\,i}^{\ \ \ bc}&=&- {\overline G}_{ia}^{(2)}\,\epsilon^{abc} \,, \nonumber \\[2mm]
	f_{+\,ij}^{\ \ \ \ c}&=&- \frac12 {\overline G}^{(4)}_{ijab}\,\epsilon^{abc}\,,  \\[2mm]
	f_{+\,ijk}&=&\frac16 {\overline G}^{(6)}_{ijkabc}\,\epsilon^{abc} \,, \nonumber \\[2mm]
	f_{+\,ij}^{\ \ \ \ k}&=& {\omega}_{ij}^{\ \ k}\,, \nonumber \\[2mm]
	f_{+\,ija}&=&- {\overline H}_{ija}\,, \nonumber \\[2mm]
	f_{+\,ia}^{\ \ \ \ b}&=&{\omega}_{ia}^{\ \ b}\, . \nonumber
\end{eqnarray}
Up to permutations of the indices (so that when all indices are lowered with the metric (\ref{eq:defeta}) the structure constants are completely antisymmetric), all the other components vanish.
Notice that the system of equations from which we derived the generalized structure constants of eq.~(\ref{strcon}) was overconstrained: this provides a non-trivial cross-check of the consistency of our results.

The above result completely defines all possible effective $d = 4$ ${\cal N} = 4$ supergravities that can be obtained in the chosen class of Type-IIA O6 compactifications with fluxes. For instance, the fermion variations and the scalar potential can be read off directly from  eqs.~(\ref{fermvar})--(\ref{WardD4}), by substituting (\ref{strcon}) and $\xi_{\alpha M}=0$. 

A similar analysis and identification of structure constants with $d=10$ fluxes was performed in \cite{dfv,Aldazabal:2008zza}, in the dual context of Type-IIB O3 compactifications.
Following the rule-of-thumb of the previous section, also in the examples of \cite{dfv,Aldazabal:2008zza} structure constants with different SO(1,1) charges appear whenever non-trivial $H$-fluxes are turned on.

\subsubsection{Jacobi identities from Bianchi identities}

Having established with eq.~(\ref{strcon}) the precise correspondence between fluxes and generalized structure constants, we can now check that the generalized Jacobi identities of eqs.~(\ref{aa})--(\ref{ee}) are in one-to-one correspondence with the Bianchi identities discussed at the end of subsection~\ref{BI}.

Since in our class of compactifications $\xi_{\alpha M} = 0$, eqs.~(\ref{aa})--(\ref{ee}) reduce just to the two constraints 
\begin{equation}
\label{qf}
	f_{\alpha R[MN} {f_{\beta PQ]}}^R=0\,,\qquad \epsilon^{\alpha\beta} f_{\alpha MNR} {f_{\beta PQ}}^R=0\,. 
\end{equation}
By taking the non-trivial components of the above constraints and substituting the explicit expressions of eq.~(\ref{strcon}), we get the following constraints on the fluxes: 
\begin{eqnarray}\label{constr}
	\left(\omega {\overline G}^{(2)}+{\overline H} {\overline G}^{(0)}\right)_{ijc}&=&0\,,\nonumber\\[2mm]
	\left(\omega {\overline G}^{(4)}+{\overline H} {\overline G}^{(2)}\right)_{ijkab}&=&0\,,\nonumber\\[2mm]
	\left(\omega \omega\right)_{ija}{}^{b}&=&0\,, \\[2mm]
	\left(\omega {\overline H}\right)_{ijka}&=&0\,,\nonumber \\
	\left(\omega {\overline H}\right)_{iabc}&=&0\,,\nonumber \\
	\left(\omega \omega\right)_{abi}{}^{k}&=&0\,.\nonumber
\end{eqnarray}
In particular,  the first four constraints in (\ref{constr}) come from the first constraint in (\ref{qf}), and the last two from the second. These are exactly the integrability conditions derived from the $d = 10$ BI in subsection~\ref{BI}. The only BI constraint that is missing is the one associated to the RR 2-form sourced by parallel D6-branes and O6-planes: this was somewhat expected, since these sources are the only ones preserving ${\cal N} = 4$ supersymmetry in four dimensions, so that their number is not constrained by the consistency of ${\cal N} = 4$ supergravity (where the number of vector multiplets is indeed a free parameter).

\subsection{$\xi \ne 0$ from the dilaton flux} 
\label{dilflux}

We elaborate here on the possibility of generating non-vanishing values for the $\xi_{\alpha M}$ parameters in the presence of a `dilaton flux', associated with an SO(1,1) axionic rescaling symmetry.  It is known that an SO(1,1) twist produces a gauging \cite{VZ5} associated with a non-vanishing $\xi$ parameter \cite{schwei}. Examples of this sort were later studied in \cite{DPP} in heterotic supergravity, we now explore the case of Type-IIA supergravity. 

The Type-IIA $d = 10$ supergravity action is invariant (at the two-derivative level) under the following SO(1,1) rescaling symmetry:
\begin{equation}
g  \to  e^{\lambda/2} \, g \,, 
\quad
B \to e^{\lambda/2} \, B \,, 
\quad
\Phi \to \Phi+\lambda \,, 
\quad
C^{(p)} \to  e^{\left(\frac{p}4-1\right)\lambda} \, C^{(p)} \,.
\end{equation}
This symmetry is a remnant of the dilatonic symmetry arising from the circle compactification of $d = 11$ supergravity. 
It still holds in the presence of localized sources, when the full action contains also the Dirac--Born--Infeld and Chern--Simons terms, as long as the world volume and the localized fields transform appropriately. 

We can then use such a symmetry to perform a duality twist. Since the metric is not invariant, such a twist corresponds also to a non-trivial Scherk--Schwarz twist, in particular to a volume non-preserving one,
\begin{equation}
{\rm tr} \ \omega\neq0 \, , 
\end{equation}
since the volume form is not invariant under dilatations. After a suitable field redefinition, however, we can go to a field basis where only the dilaton transforms non-trivially under the symmetry, and appears in the action only via derivative terms. In a such a field basis the axionic nature of this dilatonic symmetry is manifest. 

In practice, however, we can stick to the standard field basis and include an additional modification to the external derivative that takes into account the non-trivial dilaton flux:
\begin{equation}
{\cal D} = d_4 + \omega + Q \overline{\Delta} \,+ H \, \, , 
\end{equation}
where $Q$ is the charge under SO(1,1) dilatations and $\overline \Delta$ is defined by:
\begin{equation}
d\Phi=  d_4 \Phi+\overline \Delta \, .
\end{equation}
Using the generalized derivative ${\cal D}$, we can now write the BI as
\begin{equation}
{\cal D}^2 = 0 \, , 
\qquad
{\cal D} G = Q_{RR} \, .
\end{equation}
Their solutions read
\begin{eqnarray}
H&=&dB+\omega B+\frac12 \overline{\Delta} B+\overline H \,, \nonumber \\[2mm]
G^{(p+1)}&=&dC^{(p)}+\omega C^{(p)}+\frac{p-4}{4}\overline{\Delta} C^{(p)}+H\, C^{(p-2)}+\left({\overline{\bf G}e^{-B}}\right)^{(p+1)}\,,
\end{eqnarray}
and are subject to the following constraints:
\begin{eqnarray} 
\label{eq:BIconstr}
\begin{array}{c}
(d+\omega+Q\overline{\Delta}+H)^2=0 \\ \Rightarrow \\
\omega\omega=0\,, \qquad \omega\overline{\Delta}=0\,, \qquad \omega \overline H+\frac12\overline{\Delta} \overline H=0\,, \\ \\
(d+\omega+Q\overline{\Delta}+H)G^{(p+1)}=Q(\pi_{7-p}) \\ \Rightarrow \\
\omega {\overline G}^{(p+1)}+\frac{p-4}4\overline{\Delta} {\overline G}^{(p+1)}+\overline H {\overline G}^{(p-1)}=Q(\pi_{7-p})\,, \\ \\
(d+\omega+Q\overline{\Delta}+H)[\pi_{7-p}]=0 \\ \Rightarrow \\
\omega [\pi_{7-p}]+\frac{p-4}4 \overline{\Delta} [\pi_{7-p}]=0 \,, \qquad \overline H[\pi_{7-p}]=0\,.\end{array}
\end{eqnarray}
The above formulae can be easily generalized to account for localized fields and localized fluxes.

We now specialize to the case of D6/O6 brane systems. Notice that the constraints in eq.~(\ref{eq:BIconstr}) actually imply that, when $\overline{\Delta}_i\neq 0$, there must be also non-trivial metric fluxes, $\omega_{ij}{}^j$ and $ \omega_{aj}{}^j$, which in order to have ${\rm tr} \ \omega=0$ read
\begin{equation}
\omega_{ij}{}^j=\frac{3}4 \overline{\Delta}_i\,,\qquad \omega_{aj}{}^j=-\frac{3}4 \overline{\Delta}_i\,.
\end{equation}
If we now look at the covariant derivative of the universal axion we find
\begin{eqnarray}
G^{(4)}_{\mu ijk}&=&\partial_\mu C^{(3)}_{ijk}-(\omega_{ij}{}^l C^{(3)}_{lk\mu}+2{\rm Perm}_{ijk})
	-\frac12(\overline{\Delta}_i C^{(3)}_{jk\mu}+2{\rm Perm}_{ijk})\nonumber \\
	&=&\partial_\mu C^{(3)}_{ijk}+\overline{\Delta}_i C^{(3)}_{\mu jk}+2{\rm Perm}_{ijk}\,,
\end{eqnarray}
from where we can read that $\xi_{+i}=\overline{\Delta}_i$ can now be different from zero, and compute all the generalized structure constants of the ${\cal N} = 4$ gauging with a procedure similar to the one described in the previous subsections. 

Notice, however, that the generalized BI of the RR sector automatically rule out the possibility of switching on $\xi$ in the massive Type-IIA theory: indeed, the BI for $G^{(0)}$ receive only the contribution from the dilaton flux
\begin{equation}
( d_4 + \omega + Q \overline{\Delta} + H ) G^{(0)}=0 \qquad  \Rightarrow \qquad \overline{\Delta}_i {\overline G}^{(0)}=0 \, ,
\end{equation}
banning the possibility of having both these fluxes turned on at the same time 
(the only way out would be to work with D8/O8 systems, or perhaps to add 
non-geometrical/non-perturbative fluxes). The condition above
can also be identified with an ${\cal N} = 4$ Jacobi identity, in particular
with the ${}_{++i}{}^{abc}$ component of
\begin{equation}
3f_{\alpha R[MN} f_{\beta PQ]}{}^R+2\xi_{(\alpha[M}f_{\beta) NPQ]}=0 \, , 
\end{equation}
since $f_+^{abc}={\overline G}^{(0)}\epsilon^{abc}$ and for this particular component 
the first contribution in the above equation vanishes with the fluxes available in the Type-IIA theory.

The reader should keep in mind that the SO(1,1) symmetry used for the twist, both in the heterotic \cite{DPP} and in this case, is just an accidental symmetry of the two-derivative action, and does not survive as such the introduction of higher-derivative terms corresponding to $\alpha^\prime$ corrections~\footnote{We thank E.~Witten for bringing this point to our attention.}. The difficulties in finding 
explicit string constructions with non-vanishing $\xi$-parameters may be related to the analogous difficulties in 
generating non-vanishing FI terms in ${\cal N}=1$ compactifications.

\section{An ${\cal N} = 1$ family of vacua} \label{appl}
Now that we have established the connection between Type-IIA O6 flux compactifications and their consistent truncations to gauged $d= 4$, ${\cal N}=4$ supergravity, we can use the latter to study the vacuum structure of the former.
Many interesting Type-IIA vacua found recently in ${\cal N}=1$ compactifications, such as the ${\cal N}=1$ AdS$_4$ supersymmetric vacua in \cite{vzIIA,dWGKT,cfi}, and part of those in \cite{ckkltz}, are just specific truncations of the vacuum solutions of the ${\cal N}=4$ effective potential described in the previous section.
Moreover, our description could be exploited for a more systematic search for de~Sitter vacua and cosmological solutions, along the lines of \cite{cosmo}.
It might also be useful for the construction of new AdS$_4$ backgrounds dual to 3-dimensional conformal field theories with extended supersymmetry.
Finally, the extended duality group would make the study of non-geometric backgrounds more tractable.

As an example, in the following we construct and discuss the embedding in  ${\cal N}=4$ supergravity of the AdS$_4$ family of vacua found in \cite{vzIIA} and further studied in \cite{cfi,AF}.
From the ten-dimensional point of view, it corresponds to removing the ${\mathbb Z}_2\times {\mathbb Z}_2$ orbifold projection in the compactification.
We also discuss possible deformations of the solution and some properties of the dual CFT$_3$.

\subsection{${\cal N}=4$ embedding of a family of AdS$_4$ vacua} 
The family of ${\cal N}=1$ AdS$_4$ vacua found in \cite{vzIIA} corresponds to compactifications of the Type-IIA theory with O6 orientifold over ${\mathbb T}^6/{\mathbb Z}_2\times {\mathbb Z}_2$, with D6-branes and in the presence of a particular combination of RR, NSNS and geometric fluxes.
The orbifold projection implies a factorization of the 6-torus into a product of three 2-tori, ${\mathbb T}^6={\mathbb T}^2\times {\mathbb T}^2\times {\mathbb T}^2$.
For the same reason, the scalar manifold for the closed string sector on this space reduces to a K\"ahler manifold,
\begin{equation}
\rm 	\frac{SU(1,1)}{U(1)}\times \frac{SO(6,6)}{SO(6)\times SO(6)}  
\ \freccia{55pt}{{\displaystyle{\mathbb Z}_2\times {\mathbb Z}_2}}\ \frac{SU(1,1)}{U(1)}\times \left[\frac{SO(2,2)}{SO(2)\times SO(2)} \right]^3 = \left[\frac{SU(1,1)}{U(1)}\right]^7 ,
\label{Kaehler}
\end{equation}
parameterized by seven complex moduli $S$, $U_{\Lambda}$ and $T_{\Lambda}$ ($\Lambda=1,2,3$).

For the sake of simplicity, we will now consider fluxes respecting the plane interchange symmetry determined by arbitrary permutations among the ${\mathbb T}^2$ factors, though we will come back to the more general case later on.
If we indicate the fluxes and the vevs of the scalar fields as 
\begin{eqnarray}
	&& \omega_1=\frac1{3!}\omega_{ij}{}^{k}\epsilon^{ijl}\delta_{lk}\,, \quad \omega_2=\frac1{3!}\omega_{ab}{}^{k}\epsilon^{abl}\delta_{lk}\,, \quad \omega_3=\frac1{3!}\omega_{ib}{}^{c}\epsilon^{ibd}\delta_{dc}\,, \nonumber \\[3mm]
	&& {\overline H}_{0}=\frac1{3!}{\overline H}_{abc}\epsilon^{abc}\,, \quad {\overline H}_{1}=\frac1{3!}{\overline H}_{ija}\epsilon^{ija}\,, \nonumber \\[3mm]
	&& {\overline G}^{(0)}={\overline G}^{(0)}\,, \quad {\overline G}^{(2)}=\frac13{\overline G}^{(2)}_{ai}\delta^{ai}\,, \label{symmetricfluxes} \\[3mm]
	&& {\overline G}^{(4)}=-\frac1{3!}{\overline G}^{(4)}_{abij}\delta^{ai}\delta^{bj}\,, \quad {\overline G}^{(6)}=\frac1{3!}{\overline G}^{(6)}_{ijkabc}\delta^{ai}\delta^{bj}\delta^{ck}\,, \nonumber \\[3mm]
	&& s_0=\langle S\rangle \,,\quad u_0=\langle U_\Lambda\rangle \,,\quad t_0=\langle T_\Lambda\rangle \,, \nonumber
\end{eqnarray}
then the values of the fluxes giving the family of AdS$_4$ vacua read 
\begin{equation}
 \begin{array}{l} 
 \displaystyle	\frac19 \overline G^{(6)} = -t_0^2\,\overline G^{(2)} = \frac{t_0\,u_0}{6} \omega_1 = \frac{s_0\,t_0}{2} \omega_2= \frac{t_0\,u_0}{6} \omega_3\,, \nonumber \\[3mm]
\displaystyle
	\qquad \frac{t_0}{3} \overline G^{(4)} = \frac{t_0^3}{5} \overline G^{(0)} =-\frac{s_0}{2} \overline H_{0}=\frac{u_0}{2} \overline H_{1}\,, 
	\end{array}\label{vacuum1}
\end{equation}
which determine a five-parameter family of AdS$_4$ vacua (3 scalar vevs plus 2 flux parameters).
The BI associated to NSNS fields are automatically satisfied, while those of the RR sector can be satisfied by changing the number of D6-branes.
Notice that solutions can be found for arbitrary values of the scalar fields (up to quantization conditions coming from fluxes), so that arbitrary large compact volume (thus small $\alpha'$ corrections) and small string coupling can be easily realized.

To embed this family of vacua in a gauged ${\cal N}=4$ supergravity, we must be sure that, if D6-branes are present, they lie in directions parallel to the ${\cal N}=4$ O6-planes.
This requirement is equivalent to satisfying the BI for the RR 2-form without sources, namely
\begin{equation}
	5\,u_0^2\,\overline H_1^2=3\,s_0^2\,t_0^2\,\omega_2^2\,.
	\label{BI2form}
\end{equation}
This constraint reduces by one the number of free parameters of the vacua so that, once the values of the scalar vevs are chosen, only an overall constant on the fluxes remains free.
Accidentally, for this symmetric configuration, this condition also implies that the RR BI along the O6-planes is automatically satisfied, indicating that this family of solutions enjoys an ${\cal N}=8$ embedding.
In other words, the above set of fluxes and fields is also a solution of massive Type-IIA supergravity compactified on the same background without any sources.
We will come back to the importance of this observation later on. 

Inspection of the supersymmetry variations of the fermions, eq.~(\ref{fermvar}), provides a simple way to prove that the choice of fluxes of 
eq.~(\ref{vacuum1}), together with the condition (\ref{BI2form}), yields  supersymmetric AdS$_4$ solutions of the ${\cal N}=4$ 
supergravity theory constructed in the previous section.
This analysis also shows that, on the same vacua, supersymmetry is spontaneously broken to ${\cal N}=1$.
We are looking for vacua where all the fields are set to vanish, with the exception of the metric and of the scalar fields in the last line of eq.~(\ref{symmetricfluxes}), which take constant values: then solving the conditions for unbroken supersymmetry also implies that the equations of motion are satisfied.
This in turn implies that the vevs of the scalar fields minimize the potential $V$ in (\ref{WardD4}). 
Supersymmetric vacua are characterized by an SU(4)$_R$ direction $q^I$ and a set of scalar field vevs and fluxes (or gauge structure constants) such that $q^I$ is a null eigenvalue of the matrices $A_2^{IJ}$ and $(\overline{A}_{2A})^I{}_J$, defined in (\ref{eq:A2def}) and (\ref{eq:A2AB}) respectively.
The gravitino mass matrix $A_1^{IJ}$ (projected on the same SU(4)$_R$ direction) then tells us whether the vacuum is Minkowski or AdS.
If the spin-$\frac12$ field variations vanish in more SU(4)$_R$ independent directions, then the vacuum preserves more supersymmetries.

Since we have already worked out the relation between fluxes and gauge structure constants, we just need to identify the connection between the ${\cal N}=1$ moduli $S$, $U_\Lambda$, $T_\Lambda$  (and their vevs) and the ${\cal N}=4$ scalar fields ${\cal V}_\alpha$, ${\cal V}_{IJ}^M$, ${\cal V}_A^M$.
The coset representatives ${\cal V}$ obviously contain more scalars, which, however, were set to zero in our analysis of the supersymmetry conditions.
We checked that such a choice is consistent with the solution.
For the SU(1,1) sector of the scalar manifold (\ref{manifold}) the identification is easy, 
\begin{equation}
	{\cal V}_\alpha=\frac{1}{\sqrt{{\rm Im}\tau}}\left(
	\begin{array}{c}
		\tau \\
		1
	\end{array}
	\right) =\frac{1}{\sqrt{{\rm Re}S}}\left(
	\begin{array}{c}
		{-i\overline S} \\
		1
	\end{array}
	\right)\,. 
\end{equation}
For the SO(6,6) sector the identification is more involved.
After some calculations we find for ${\cal V}^{IJ\ M}$ 
\begin{eqnarray}
	{{{\cal V}}}^{IJ\ M}&=& \Bigl [ \delta^{M}_{\Lambda} (x_\Lambda^1 \alpha_\Lambda+{\widetilde x}_\Lambda^1 \beta_\Lambda)^{IJ}\,,\quad \delta^{M-3}_{\Lambda} (x_\Lambda^2 \alpha_\Lambda+{\widetilde x}_\Lambda^2 \beta_\Lambda)^{IJ}\,, \label{VIJM}\\
	&& \quad \delta^{M-6}_{\Lambda} (x_\Lambda^3 \alpha_\Lambda+{\widetilde x}_\Lambda^3 \beta_\Lambda)^{IJ}\,, \quad \delta^{M-9}_{\Lambda} (x_\Lambda^4 \alpha_\Lambda+{\widetilde x}_\Lambda^4 \beta_\Lambda)^{IJ}\Bigr] \,,\nonumber
\end{eqnarray}
where $\alpha_{\Lambda}$ and $\beta_{\Lambda}$ are six four-by-four matrices that map SU(4) indices into SO(6),
\begin{equation}
	\alpha_1 = \frac{i}{2} \sigma^2 \otimes \sigma^1 \, ,  
	\qquad
	\alpha_2 = - \frac{i}{2} \sigma^2 \otimes \sigma^3 \, , 
	\qquad
	\alpha_3 = \frac{i}{2} {\bf 1}_2 \otimes \sigma^2 \, , 
	\label{alphamat}
\end{equation}
\smallskip
\begin{equation}
	\beta_1 = -\frac{1}{2} \sigma^1 \otimes \sigma^2 \, , 
	\qquad
	\beta_2 = -\frac{1}{2} \sigma^2 \otimes {\bf 1}_2 \, , 
	\qquad
	\beta_3 = \frac{1}{2} \sigma^3 \otimes \sigma^2 \, , 
	\label{betamat}
\end{equation}
and
\begin{equation}
\left(
\begin{array}{c}
x_\Lambda^1+i\,{\widetilde x}_\Lambda^1 \\
x_\Lambda^2+i\,{\widetilde x}_\Lambda^2 \\
x_\Lambda^3+i\,{\widetilde x}_\Lambda^3 \\
x_\Lambda^4+i\,{\widetilde x}_\Lambda^4
\end{array}
\right) =
\sqrt{\frac{2}{Y_\Lambda}}\left(
\begin{array}{c}
1 \\
U_\Lambda T_\Lambda \\
iU_\Lambda \\
iT_\Lambda
\end{array}
\right)\,, \qquad {\rm with}\quad Y_\Lambda=(T_\Lambda+{\overline T}_\Lambda)(U_\Lambda+{\overline U}_\Lambda)\,.
\end{equation}
Analogously, for ${\cal W}^{M IJ}={{\cal V}}^M_A Q^{A\,IJ}$, where $Q^A = \{\alpha_\Lambda,\beta_\Lambda\}$, we can find a similar expression to the one in (\ref{VIJM}), but with different scalar functions ($y_\Lambda$ instead of $x_\Lambda$): 
\begin{equation}
\left(
\begin{array}{c}
y_\Lambda^1+i\,{\widetilde y}_\Lambda^1 \\
y_\Lambda^2+i\,{\widetilde y}_\Lambda^2 \\
y_\Lambda^3+i\,{\widetilde y}_\Lambda^3 \\
y_\Lambda^4+i\,{\widetilde y}_\Lambda^4
\end{array}
\right) =
\sqrt{\frac{2}{Y_\Lambda}}\left(
\begin{array}{c}
1 \\
-U_\Lambda {\overline T}_\Lambda\\
iU_\Lambda \\
-i{\overline T}_\Lambda 
\end{array}
\right)\,, 
\end{equation}
which corresponds to the exchange of $T_\Lambda$ with $-{\overline T}_\Lambda$ (or $U_\Lambda$ with $-{\overline U}_\Lambda$ if the complex conjugate is taken) in the expressions for the $x_\Lambda$.
It is easy to check that, with this choice of parameterization, the constraints (\ref{pseudoreality}) and (\ref{cosetconstraint}) are satisfied and the known ${\cal N}=1$ results in the truncated limit can be recovered.
This last check can be performed by looking at the gravitino mass matrix.
In the basis for the $(\alpha_\Lambda,\beta_\Lambda)$ matrices of eqs.~(\ref{alphamat})--(\ref{betamat}), the gravitino mass matrix is diagonal, with three degenerate eigenvalues (due to the plane interchange symmetry of the fluxes).
The fourth eigenvalue is the one surviving the orbifold projection and 
after using eq.~(\ref{strcon}) reads
\begin{eqnarray}
	A_1^{44}&\propto& \frac{e^{K/2}}{2}\left[ G^{(6)}+i G^{(4)} (T_1+T_2+T_3)
		-G^{(2)}(T_1T_2+T_2 T_3+T_3T_1)-iG^{(0)} T_1 T_2 T_3 \right. \nonumber \\
	&& \left. i H_0 S-i H_1 (U_1+U_2+U_3)+\omega_1(T_1 U_1+T_2 U_2+T_3 U_3)-\omega_2 S(T_1+T_2+T_3)\right.\nonumber \\
	&& \left.-\omega_3(T_1 U_2+T_1 U_3+T_2 U_1+T_2 U_3+T_3 U_1+T_3 U_2) \right]\,, 
\label{superpotential}
\end{eqnarray}
which nicely matches the expression of the ${\cal N}=1$ superpotential found in \cite{dkpz,vzIIA}.

Using the same conventions, the SU(4)$_R$ direction corresponding to preserved supersymmetry is thus 
\begin{equation}
	q_I=\delta_I^4\,,
\end{equation} 
i.e.~the one preserved by the orbifold projection.
It is rather easy now to check explicitly that the fermion supersymmetry variations projected along this direction vanish precisely when the AdS$_4$ constraints (\ref{vacuum1})--(\ref{BI2form}) on the fluxes and the field vevs are satisfied.
One way to do so without doing any computation is to notice that, once the $A_{(2)}$ matrices entering the spin-$\frac12$ supersymmetry transformations are contracted with the SU(4)$_R$ vector $q_I$, they reconstruct the ${\cal N}=1$ F-terms.
The vanishing of the latter then ensures the vanishing of the ${\cal N}=4$ fermion variation.
Notice that, because of the particular form of the K\"ahler manifold (\ref{Kaehler}) and of the flux superpotential (\ref{superpotential}), the ${\cal N}=1$ F-terms read: 
\begin{eqnarray}
	F_S&=&\left.e^{K/2}W\right|_{S\to-{\overline S}} \,, \label{FS} \\[2mm]
	F_{U_\Lambda}&=&\left.e^{K/2}W\right|_{U_\Lambda\,\to\,-{\overline U}_\Lambda}\,, \label{FU}\\[2mm]
	F_{T_\Lambda}&=&\left.e^{K/2}W\right|_{T_\Lambda\,\to\,-{\overline T}_\Lambda}\,. \label{FT}
\end{eqnarray}
These conditions exactly match the relation between the ${\cal N}=4$ fermion variation $A_{(2)}$ and the gravitino mass $A_{(1)}$: the dilatino variation $A_{(2)}{}_I{}^J$ has indeed the same expression of $A_{(1)}$ with the substitution of ${\cal V}_\alpha$ with ${\cal V}^*_\alpha$ which corresponds to eq.~(\ref{FS}), while the components $\epsilon_{HKL} A_{(2)}{}^{HK}{}_I{}^J$ and $A_{(2)}{}^{L 4}{}_I{}^J$ correspond to substitute in $A_{(1)}$ one ${\cal V}^M{}^{IJ}$ with ${\cal W}^M{}^{IJ}$, thus exactly to the substitutions in eqs.~(\ref{FU}) and (\ref{FT}). 

We can also check that the direction $q^I=\delta^I_4$ is indeed the only one that annihilates the fermion variation.
This means that even when the orbifold is removed we have ${\cal N}=1$ AdS$_4$ vacua, this time arising via spontaneous symmetry breaking from ${\cal N}=4$. 

As we have discussed at length in the previous section, the reduction from 10 to 4 dimensions with fluxes leads to an ${\cal N} = 4$ \emph{gauged} supergravity.
This implies that the choice of fluxes (\ref{vacuum1}), leading to the family of AdS$_4$ vacua presented in \cite{vzIIA}, corresponds to a non-trivial gauge group, specified by (\ref{strcon}).
More details on the general structure of the gauge group and its symplectic embedding can be found in the Appendix.
It is interesting, however, to point out that the general gauge group reduces to the semidirect product of SU(2) with the group $N_{9,3}$ associated to a 3-step nilpotent algebra:
\begin{equation}
	G = {\rm SU}(2) \rtimes N_{9,3}.
\end{equation}
More in detail, we can summarize the gauge algebra specified by the choices (\ref{vacuum1}) and (\ref{BI2form}) as
\begin{equation}
	[X_i, X_j] = \epsilon_{ijk} X_k, \qquad [X_i,A^I_j] = \epsilon_{ijk} A^I_k, 
\end{equation}
\begin{equation}
	[A^1_i,A^1_j] = \epsilon_{ijk} A^2_k, \qquad [A^1_i, A^2_j] = \epsilon_{ijk} A^3_k.
\end{equation}
Here $X_i$ are the SU(2) generators and $A_i^I \in {\mathfrak n}_{9,3}$, for $I=1,2,3$. 
At the ${\cal N} = 1$ critical point the 9 vectors gauging the nilpotent group are massive and the surviving gauge group is 
\begin{equation}
	G_{vac} = {\rm SU}(2).
\end{equation}
We point out that this gauge group, however, does not match the full symmetry group of the corresponding type IIA solution.
We will see in the next section that the $d= 10$ background has an SU(2)$^3$ isometry group and that the Scherk--Schwarz reduction sees only its truncation to $G_{vac} = {\rm SU}(2)$.
As we already explained, all BI are satisfied without source terms.
However, the presence of O6-planes from the orientifold projection requires the further presence of 16 D6-branes (and their images) to cancel the corresponding charge: we can do this by placing the D6-branes on top of the O6-planes so that their charge and tension cancel locally.
This configuration allows to solve the $d= 10$ equations of motion and BI exactly, without the need of smearing the sources.
This implies that at the ${\cal N} = 1$ vacuum there are also matter fields associated to the fluctuations of the D6-branes, which we put to zero to find the vacuum solution.
In particular there are 8 O6-planes and 2 D6-branes on top of each O-plane to cancel their charge and tension.
This configuration adds an extra SO(4)$^8$ gauge factor to the $d = 4$ effective action.
If we are interested in recovering the full ${\cal N}=4$ effective theory around this vacuum, we should in principle consider also these fields, which enlarge both the scalar manifold and the gauge group. 
We can anticipate that many of the extra scalar fields will get mass from fluxes.

Since the D6 and O6 charges cancel without the need of a net flux contribution, the solution will survive also in the absence of the orientifold projection.
The family of AdS$_4$ solutions described above is then also a solution of the massive oriented Type-IIA equations of motion.
The cancellation of the D6-brane charge is also a signal that the truncated $d = 4$ theory without the orientifold projection can be embedded in a gauged ${\cal N}=8$ supergravity.
Indeed, as shown in the Appendix, the gauge algebra can be embedded in ${\mathfrak e}_{7(7)}$.
In this context we can discuss again the structure of the effective theory and the moduli stabilization process.
While leaving all the technical details for the Appendix, we can summarize here a couple of interesting results of this analysis.

The gauge group of the resulting ${\cal N} = 8$ truncation is also a semidirect product of a compact group, in this case SU(2) $\times$ U(1), with a nilpotent group, now of dimension 24.
On the vacuum, all the vector fields associated to the nilpotent group become massive as they should.
The compact part has an interesting structure, because the U(1) group is compatible with the R-symmetry group of a residual ${\cal N} = 2$ supersymmetric theory.

Gauged maximal supergravities in $d=4$ have a natural link with M-theory reductions.
While most of the massive IIA fluxes are perturbative also from the M-theory point of view, being either 4- and 6-form fluxes or metric fluxes, the $\overline{G}^{(0)}$ flux has clearly a non-perturbative origin.
This can be explicitly seen from the embedding of our reduced model in ${\cal N} = 8$ supergravity and the attempt at interpreting this theory as a Scherk--Schwarz reduction of M-theory.
The $\overline{G}^{(0)}$ flux induces a gauging that involves the vector field coming from the dual metric along the M-theory/IIA circle, therefore it cannot be obtained in a usual compactification scheme.
According to ref.~\cite{Hull:1998vy}, the massive IIA theory would arise from M-theory by compactifying on a collapsing twisted 3-torus (in other words, by taking a suitable zero-size limit of a compactification on ${\mathbb T}^3$ with metric flux $\omega_{mn}{}^p$). 
This picture nicely agrees with our analysis of the ${\cal N} = 8$, $d=4$ gauged supergravity: $\overline G^{(0)}$ induces a gauging involving the vector fields $C^{(7)}_{\mu mnqrst}$, $B_{\mu m}$  and $B_{\mu t}$ (where the index $m$ is along the twisted 3-torus, while $t$ is not).
After the M-theory uplift these vectors are mapped into $A^{(6)}_{\mu pqrst}$, $V_\mu^m$ and $A^{(3)}_{\mu tp}$, which are indeed gauged by the metric flux on the 3-torus (see also (\ref{Zmd})--(\ref{Zmu}) in the Appendix).

In view of our analysis, this correspondence can be pushed further, extending it from fluxes to sources.
As already stated, $\overline G^{(0)}$ gauges the vector of the dual metric, which couples electrically to KK6-monopoles.
This suggests that M-theory KK6-monopoles are related to D8-branes, i.e.~the sources of the IIA mass parameter.
The above connection can be described by the following chain of dualities:
\begin{equation} 
 	\begin{array}{ccccccc}
		\rm	IIA & & \rm IIB & & \rm IIA && \rm M \\[2mm]
		\overline G^{(0)} & \freccia{35pt}{\displaystyle {\rm T}_{m\phantom{p}}} &\overline G^{(1)}_m &\freccia{35pt}{\displaystyle {\rm T}_{n\phantom{p}}}&\overline G^{(2)}_{mn}&\freccia{35pt}{\displaystyle S^1_p}& \omega_{mn}{}^p \\[2mm]
 [\pi_8]_q & \freccia{35pt}{\displaystyle {\rm T}_{m\phantom{p}}} & [\pi_7]_{qm} &\freccia{35pt}{\displaystyle {\rm T}_{n\phantom{p}}}&[\pi_6]_{qmn}&\freccia{35pt}{\displaystyle S^1_p}& [\kappa_6]_{qmn}{}^p \\[3mm]
	\end{array}\,.
\end{equation}
In the above scheme, T$_m$ and T$_n$ denote T-dualities along the $m$ and $n$ directions ($m \neq n$), $S^1_p$ the M-theory uplift.
Similarly, $[\pi_8]$, $[\pi_7]$, $[\pi_6]$ and $[\kappa_6]$ denote the Poincar\'e duals of the D8-, D7-, D6-brane world-volumes and of the M-theory KK6-monopole, respectively.
Thus D8-branes would correspond to M-theory KK6-monopoles localized on the twisted 3-torus, with the fibres of the KK6-monopole and of the twisted 3-torus identified.

\subsection{The geometry of the massive IIA vacuum}

We now discuss the geometry of the $d= 10$ solution.
In \cite{AF} it was shown that, in the case $t_1=t_2=t_3$, the ${\cal N}=1$ AdS$_4$ vacua of eq.~(\ref{vacuum1}) correspond to compactifications on 
$AdS_4\times X_6$, with the internal manifold $X_6$ having the topology of $(S_3\times S_3)/{\mathbb Z}_2^3$, 
where the $S_3$ were produced by the geometric fluxes and the ${\mathbb Z}_2^3$ projection was due to the ${\mathbb Z}_2\times {\mathbb Z}_2$ orbifold plus the O6 orientifold involution.
We now show that, even in the generic case, the solution of our ${\cal N}=4$ gauged supergravity theory corresponds to a compactification on a 
$S_3\times S_3$ manifold  with RR and NSNS fluxes turned on and an O6 orientifold involution that exchanges the two 3-spheres.
We discuss the geometric structure of the internal manifold, showing explicitly that it solves the full massive IIA equations 
even for generic fluxes not satisfying the plane-interchange symmetry of (\ref{symmetricfluxes}) and (\ref{vacuum1}).
This analysis, which follows the lines of the analogous discussion in \cite{AF}, will also lead us to the correct identification of the flux quantization conditions as well as of the possible deformations of our background.

A Scherk--Schwarz reduction is equivalent to a compactification on a local group manifold, which goes under the name of twisted torus.
In our case, the metric on the internal 6-manifold $Y_6$ can be written as 
\begin{equation}
	ds_{Y_6}^2=\sum_{\Lambda=1}^{3} \frac{t_\Lambda}{{\widehat u}_\Lambda} (\eta^\Lambda)^2 + t_\Lambda{\widehat u}_\Lambda ({\widetilde \eta}^\Lambda)^2,
\end{equation}
in terms of two sets of three globally defined twisted-torus 1-forms $(\eta^\Lambda,{\widetilde \eta}^\Lambda)=(\eta^i,\eta^a)$ that satisfy the conditions 
\begin{equation}
	\begin{array}{rcl}
	d\eta^\Lambda&=&\omega_1^\Lambda \,\eta^\Sigma \eta^\Gamma+\omega_2^\Lambda\, {\widetilde \eta}^\Sigma {\widetilde \eta}^\Gamma \,,  \\[2mm]
	d{\widetilde \eta}^\Lambda&=&\omega_{3 \Sigma \Gamma} \,{\eta}^\Sigma {\widetilde \eta}^\Gamma+\omega_{3 \Gamma \Sigma} \,{\widetilde \eta}^\Sigma {\eta}^\Gamma \,, 
	\end{array}
\end{equation}
where $(\Lambda,\Sigma,\Gamma)=(1,2,3)$ and cyclic permutations. 
We recall here that $t_\Lambda \equiv {\rm Re}\, T_\Lambda$ are the volume moduli of the three ${\mathbb T}^2$'s before twisting and that $\widehat u_\Lambda$ are related to the ${\cal N} = 1$ subsector (\ref{Kaehler}) of the moduli space (\ref{manifold}) by
\begin{equation}
	{\rm Re}\,S = {e}^{-\Phi}	\sqrt{\frac{ t_1 t_2 t_3}{\widehat{u}_1\widehat{u}_2\widehat{u}_3}}\,, \qquad 
	{\rm Re}\,U_\Lambda = {e}^{-\Phi}\sqrt{\frac{t_1 t_2 t_3\widehat{u}_\Sigma \widehat{u}_\Gamma}{\widehat{u}_\Lambda}}\,. 
\end{equation}

On a generic ${\cal N}=1$ vacuum, these moduli satisfy 
\begin{equation}
	\frac{3}{\widehat{u}_\Sigma \widehat{u}_\Gamma} = \frac{\omega_1^\Lambda}{\omega_2^\Lambda}\,, \qquad 
	\frac{\widehat{u}_\Lambda}{\widehat{u}_\Gamma} = \frac{\omega_1^\Lambda}{\omega_{3 \Sigma \Gamma}}\,,\qquad \frac{\widehat{u}^\Lambda t^\Sigma}{t^\Lambda \widehat{u}^\Sigma} = \frac{\omega_1^\Lambda}{\omega_1^\Sigma}\,,
	\label{vacuum2}
\end{equation} 
where it is now clear that we did not impose the plane interchange symmetry leading to (\ref{vacuum1}).
We can now show that the space resulting from imposing (\ref{vacuum2}) is the product of two 3-spheres.
To do so, it is useful to change basis and use another set of vielbeins, ($\xi^\Lambda,{\widetilde \xi}^{\Lambda}$), defined as 
\begin{equation}
	\begin{array}{rcl}
	\xi^\Lambda&\equiv&\displaystyle \sqrt{\omega_1^\Sigma \omega_1^\Gamma}\left(\eta^\Lambda+ \frac{\widehat{u}_\Lambda}{\sqrt{3}}\, \widetilde \eta^\Lambda\right),\nonumber\\[5mm]
	\widetilde \xi^\Lambda&\equiv& \displaystyle\sqrt{\omega_1^\Sigma \omega_1^\Gamma}\left(\eta^\Lambda- \frac{\widehat{u}_\Lambda}{\sqrt{3}}\, \widetilde \eta^\Lambda\right).
	\end{array}
\end{equation}
These new vielbeins satisfy the simple conditions 
\begin{equation}
	\begin{array}{rcl}
	d\xi^\Lambda=\xi^\Sigma \xi^\Gamma\,, \nonumber \\[3mm]
	d{\widetilde \xi}^\Lambda={\widetilde \xi}^\Sigma {\widetilde \xi}^\Gamma\,, 
	\end{array}
\end{equation}
corresponding to a realization of an SU(2) $\times$ SU(2) group manifold, namely the product of two 3-spheres.
It should be noted that just like the $(\eta^\Lambda,\widetilde \eta^\Lambda)$ vielbeins of the original basis, also the ($\xi^\Lambda,{\widetilde \xi}^{\Lambda}$) vielbeins are globally defined, because $S^3$ is a  parallelizable manifold.

In this new basis the metric takes the simple form 
\begin{equation}
	ds_{Y_6}^2= \rho^2 \left( (\xi^\Lambda)^2+({\widetilde \xi}^\Lambda)^2-\xi^\Lambda {\widetilde \xi}^\Lambda \right),
	\label{metrbas}
\end{equation}
with the overall radius given by
\begin{equation}
	\rho \equiv \left(\frac{t_1 t_2 t_3}{\left({\omega_1^1} {\omega_1^2} {\omega_1^3}\right)^2 \widehat{u}_1\widehat{u}_2\widehat{u}_3}\right)^{1/6}.
\end{equation}
The metric is actually that of two $S^3$ at angle.
Since the angle reduces the SO(4)$^2$ isometry of the two spheres to SU(2)$^3$,  the internal manifold corresponds to the coset 
\begin{equation}
	Y_6 = \frac{{\rm SU}(2) \times {\rm SU}(2) \times{\rm SU}(2)}{{\rm SU}(2)}.
\end{equation}
Once more we can see that the full symmetry group of this background, namely SU(2)$^3$, is larger than the one we see at the vacuum of our $d= 4$ gauged supergravity model, which is just SU(2).
The reason for this lies in the fact that the gauged supergravity model of the previous section is obtained by performing a Scherk--Schwarz reduction on the two $S^3$ at angle.
Each $S^3$ has a metric that is invariant under SU(2)$_L \times$ SU(2)$_R$, where the $L,R$ subscript refers to left or right multiplication by the SU(2) group.
Because of the angle, the metric (\ref{metrbas}) is invariant only under SU(2)$_{1,L} \times$ SU(2)$_{2,L} \times$ SU(2)$_{D,R}$, where the subscripts $1,2$ refer to the two spheres and SU(2)$_{D,R}$ is the diagonal right action.
The Scherk--Schwarz reduction, however, keeps only modes that are singlets under the action from the left of the isometry group of the internal local group manifold.
This means that only left invariant Killing vectors will survive and hence only the SU(2)$_{D,R}$ isometry group can be seen in the reduced theory.

Note that, out of the various parameters that control the vacua, only the combination corresponding to the total volume enters the metric.
We can actually show that this is also related to the ratio of two quantized parameters, which control all the other quantities characterizing our solution.
Using the relation between fluxes and moduli of eq.~(\ref{vacuum2}), we can rewrite the AdS$_4$ solution in the $\xi$ basis as a function of two integers: 
$g_0$ and $g_6$.
The metric, the dilaton and the fluxes then read
\begin{eqnarray}
	ds^2_{IIA}&=& ds_{AdS_4}^2+\rho^2 \left( (\xi^\Lambda)^2+({\widetilde \xi}^\Lambda)^2-\xi^\Lambda {\widetilde \xi}^\Lambda \right) \,, \qquad \rho^2=\frac{5^{1/6}}{2^{2/3}} \left(\frac{g_6}{g_0}\right)^{1/3}\, \,, \nonumber \\[2mm]
	e^{-2\Phi}&=&\frac{2^{4/3}\cdot3}{5^{5/6}}\, (g_0^5 \, g_6)^{1/3}\,, \nonumber \\[3mm]
	G^{(0)} &=& g_0\,, \nonumber \\[3mm]
	G^{(2)} &=& -\frac{(g_0^2\, g_6)^{1/3}}{2^{5/3} \cdot 5^{1/3}}\,\left(\xi^1 {\widetilde \xi}^1+\xi^2 {\widetilde \xi}^2+\xi^3 {\widetilde \xi}^3 \right)\,,  \\[3mm]
	G^{(4)} &=& \frac{9(g_0\, g_6^2)^{1/3}}{2^{10/3}\cdot5^{2/3}}\,\left(\xi^2 {\widetilde \xi}^2 \xi^3 {\widetilde \xi}^3+ \xi^3 {\widetilde \xi}^3 \xi^1 {\widetilde \xi}^1+\xi^1 {\widetilde \xi}^1 \xi^2 {\widetilde \xi}^2\right)\,, \nonumber \\[3mm]
	G^{(6)} &=& \frac{3^3}{2^5} g_6 \,\xi^1 {\widetilde \xi}^1 \xi^2 {\widetilde \xi}^2 \xi^3 {\widetilde \xi}^3 \,, \nonumber \\[3mm]
	H &=& \frac{1}{2^{5/3} \cdot 5^{1/3}}\, \left(\frac{g_6}{g_0}\right)^{1/3}\,\left({\widetilde \xi}^1\xi^2\xi^3-\xi^1 {\widetilde \xi}^2{\widetilde \xi}^3 +{\widetilde \xi}^2\xi^3\xi^1-\xi^2 {\widetilde \xi}^3{\widetilde \xi}^1 +{\widetilde \xi}^3\xi^1\xi^2-\xi^3 {\widetilde \xi}^1{\widetilde \xi}^2\right)\,. \nonumber
\end{eqnarray}
It should be noted that $G^{(4)}$ and $H$ are trivial in cohomology on the spheres.
This means that to generate the background above we really need to switch on only two non-trivial fluxes~\footnote{Notice that flux quantization has to be imposed on the combinations $(Ge^{B})^{(n)}$, which are closed because of the BI (see eq.~(\ref{eq:RRBI})). In our conventions this implies that the quantized fluxes are the ${\overline G}^{(n)}$ instead of the $G^{(n)}$.}:
\begin{equation}
	{\overline G}^{(0)} = g_0\,, \qquad
	{\overline G}^{(6)} = g_6\,\xi^1 {\widetilde \xi}^1 \xi^2 {\widetilde \xi}^2 \xi^3 {\widetilde \xi}^3 \,.
\end{equation}
All the other fluxes are trivial, because $H=dB$, $G^{(2)}=-B G^{(0)}$ and $G^{(4)}=dC^{(3)}+\frac12 B B G^{(0)}$, with
\begin{eqnarray}
	B &=& \frac{1}{2^{5/3} \cdot 5^{1/3}}\,\left(\frac{g_6}{g_0}\right)^{1/3}\,\left(\xi^1 {\widetilde \xi}^1+\xi^2 {\widetilde \xi}^2+\xi^3 {\widetilde \xi}^3\right)\,,  \\[2mm]
	C^{(3)} &=& -\frac{4}{2^{4/3}\cdot5^{2/3}} \left(g_0 g_6^2\right)^{1/3}\,\left({\widetilde \xi}^1\xi^2\xi^3+\xi^1 {\widetilde \xi}^2{\widetilde \xi}^3 +{\widetilde \xi}^2\xi^3\xi^1+\xi^2 {\widetilde \xi}^3{\widetilde \xi}^1 +{\widetilde \xi}^3\xi^1\xi^2+\xi^3 {\widetilde \xi}^1{\widetilde \xi}^2\right)\,.\nonumber
\end{eqnarray}

Since this solution preserves ${\cal N} = 1$ supersymmetry, we can see that the fluxes and the geometry satisfy the SU(3) group-structure constraints derived in \cite{Lust:2004ig}.
We recall that in the case of a Scherk--Schwarz reduction, the internal manifold always defines a trivial group structure.
Each supersymmetry will especially define a complex structure, with its associated 2-form $J$, and a holomorphic 3-form $\Omega$.
Given these forms, the fluxes will obey the supersymmetry constraints derived in \cite{Lust:2004ig}, which, in the string frame and with the warp factor set to 1, read
\begin{equation}
	\begin{array}{c}
		\vspace{.4cm} \displaystyle dJ = 2\widetilde m {\rm Re} \Omega\ ,\qquad d \Omega= i\left(W_2^- \, J -\frac43 \widetilde m J^2\right) ,\qquad H= -2m {\rm Re} \Omega ;\\
		\displaystyle G^{(0)}=5m{\rm e}^{-\Phi}\,,\quad {\rm e}^{\Phi} G^{(2)}=-W_2^-+\frac13 \widetilde m J ,\quad G^{(4)} = \frac32 m {\rm e}^{-\Phi} J^2 ,\quad  G^{(6)}=-\frac12\widetilde m {\rm e}^{-\Phi}J^3.
	\end{array}
\end{equation}
The solution is given by the SU(3) structure defined by 
\begin{equation}
	J = \frac{3^{1/2}\cdot 5^{1/6}}{2^{5/3}}\,\left(\frac{g_6}{g_0}\right)^{1/3} \, \left(\xi^1 \, \widetilde \xi^1+\xi^2 \, \widetilde \xi^2+\xi^3 \, \widetilde \xi^3\right) 
\end{equation}
and the (3,0)-form
\begin{equation}
		\Omega = \frac{5^{1/4}}{2^3}\sqrt{\frac{g_6}{g_0}} \left ({\widetilde \xi}^1 -e^{2\pi i/3}\xi^1\right)\,
		\left ( {\widetilde \xi}^2 -e^{2\pi i/3}\xi^2\right)\, \left ({\widetilde \xi}^3 -e^{2\pi i/3}\xi^3\right)\,.
\end{equation}
The other parameters are 
\begin{equation}
	W_2^- = 0, \qquad \widetilde m = -\sqrt{15}\, m =- \frac{1}{2^{2/3}\cdot 5^{1/12}}\left(\frac{g_0}{g_6}\right)^{1/6}.
\end{equation}
This shows that the metric of $Y_6$, leading to our ${\cal N} = 4$ supergravity vacuum, is actually nearly-K\"ahler.
It therefore coincides with one of the special massive IIA AdS$_4$ solutions found in \cite{Behrndt:2004km}.

\FIGURE[t]{
\psfrag{a0}{$T^6/{\mathbb Z}_2$}
\psfrag{b0}{\hspace{-5pt}${\cal N}=4,8$}
\psfrag{c0}{$g_0=0$} 
\psfrag{a1}{\cite{vzIIA,dWGKT,cfi}} 
\psfrag{b1}{\cite{Behrndt:2004km,vzIIA,cfi}}
\psfrag{c1}{\cite{Acharya:2003ii,vzIIA,cfi}} 
\psfrag{a2}{$\displaystyle\frac{{\widetilde m}^2}{m^2}=0$}
\psfrag{b2}{$\displaystyle\frac{{\widetilde m}^2}{m^2}=15$} 
\psfrag{c2}{$\displaystyle\frac{{\widetilde m}^2}{m^2}=\infty$}
\epsfig{file=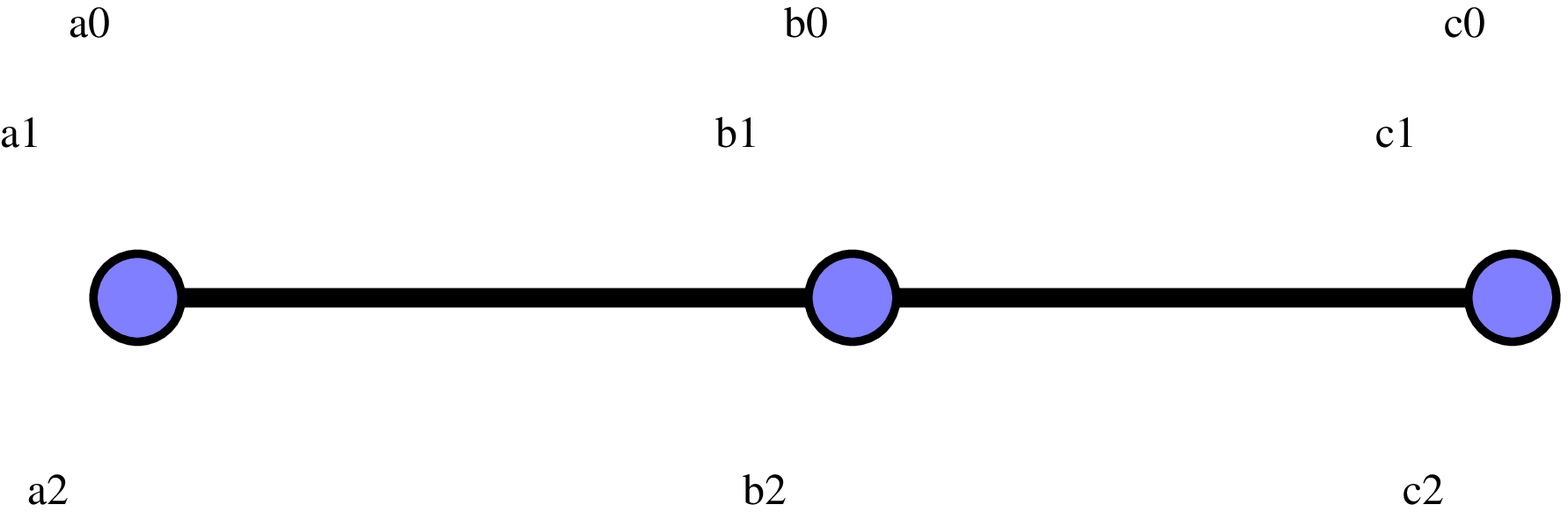,width=0.65\textwidth}
\caption{The family of AdS$_4$ solutions discussed in the text. When ${\widetilde m}=0$ there are no metric fluxes,
the geometry collapses to $T^6/{\mathbb Z}_2$ O6 orientifold. As $\frac{{\widetilde m}^2}{m^2}\neq0$ metric fluxes deform
the torus into $S^3\times S^3$, when $\frac{{\widetilde m}^2}{m^2}=15$ the net D6-brane charges cancel and
the solution allows a description in terms of ${\cal N}=4$ (or ${\cal N}=8$ in the absence of O6-planes) gauged supergravity.
In the limit $m^2=0$ the massive parameter vanishes and the solution admit a geometrical M-theory uplift.}
}

As noted in \cite{AF}, we could still solve the supersymmetry conditions by adding smeared D6-branes that modify the 2-form BI and hence relax the relation between the parameters $m$ and $\widetilde m$. 
For $\widetilde m^2 > 15 m^2$ one can obtain new solutions by adding D6-branes, because the 2-form BI reduces to
\begin{equation} \label{eq:BIwithsources}
	d G^{(2)} + H G^{(0)} = \frac23 {\rm e}^{-\Phi} \left(\widetilde m^2 - 15 m^2\right) {\rm Re}\Omega = Q(\pi_6).
\end{equation}
From the flux point of view, this means that we can introduce a further parameter corresponding to the D6-brane density, which allows to interpolate between the cases with $\overline G^{(0)} = 0$, $\overline G^{(6)} \neq 0$ of \cite{Acharya:2003ii}, the one with both $\overline G^{(0)} \neq 0$ and $\overline G^{(6)} \neq 0$ and ${\widetilde m}^2=15 m^2$ of \cite{Behrndt:2004km}, its generalizations (with $\overline G^{(0)} \neq 0$ and $\overline G^{(6)} \neq 0$ and ${\widetilde m}^2\neq 15 m^2$), and  finally the case $\overline G^{(0)} \neq 0$, $\overline G^{(6)} = 0$. 
The latter case corresponds to switching off the metric fluxes and the geometry becomes ${\mathbb T}^6/{\mathbb Z}_2$, corresponding to the unorbifolded version of the solutions
of \cite{vzIIA,dWGKT,Acharya:2006ne}. 
The case where the massive parameter is vanishing is especially interesting, because it allows for a lift to M-theory, where the resulting space should have $G_2$ holonomy. The $S^3 \times S^3$ manifold can actually be used as the base of a non-compact $G_2$-holonomy manifold built from its cone \cite{Atiyah:2001qf}, and the relation between this cone and the IIA solution has been discussed in \cite{Acharya:2003ii}.

\subsection{Scales} 
\label{sec:scales}

As discussed above, in the absence of a net D6-brane charge, the solutions can be parameterized by two integer numbers: $g_6$ and $g_0$.
Neglecting for the moment order one coefficients, the scaling of the volume and the dilaton with respect to those parameters reads
\begin{equation}
\rho^2\sim \left(\frac{g_6}{g_0}\right)^{1/3}\,,\qquad e^{2\Phi}\sim \frac{1}{g_0^{5/3}g_6^{1/3}}\sim \frac{1}{g_0^2 \rho^2}\,.
\end{equation}
It is easy to see that for $g_6\gg g_0$ both the volume and the inverse string coupling can be made arbitrary large, so as to justify the classical supergravity calculation.

We need now to check whether the AdS$_4$ scale (which gives the scale of the massive modes) can be made parametrically smaller than the KK scale, to permit a 4d effective field theory description.
The KK scale is set by the radius of the spheres $\rho$, while the AdS$_4$ length can be extracted by the 4d Hubble parameter
\begin{equation}
\ell_{AdS}^{-2}=\frac{V_0}{M_P^2}\sim \frac{g_0^2\sqrt{g}}{e^{-2\Phi}\sqrt{g}}\sim \rho^{-2}\,,
\end{equation}
where $V_0$ is the vev of the $d = 4$ potential and $M_P$ is the $d = 4$ Planck mass.
We can see that independently of the value of the parameters in this case the AdS scale is always of the same order of the KK scale. 
This is a common feature of this type of compactifications (as in $AdS_5\times S^5$, $AdS_4\times S^7$, etc.), where the positive energy contributions from the RR and NSNS fluxes to the effective potential are compensated by the negative contribution from the geometric fluxes, i.e.~the curvature of the internal manifold; therefore the net contribution to the $d = 4$ curvature is basically given by the internal curvature itself, giving the relation between the KK scale and the AdS length.

The relation between the AdS length and the KK scale also implies that, for this class of solutions, gauged supergravity around the vacuum does not coincide with the full $d = 4$ effective field theory. 
Rather it represents just a particular truncation,  describing a subset of the higher-dimensional spectrum in terms of a $d = 4$ gauged supergravity. 
The latter can thus be seen as a tool for generating solutions.
This explains why for example the $d = 4$ gauged supergravity sees only an SU(2) gauged group instead of the expected SU(2)$^3$ associated to the full isometry of the solution.
The Scherk--Schwarz reduction procedure truncated away part of the massless spectrum and kept part of the KK modes in order to reconstruct a Lagrangian consistent with the ${\cal N} =4$ and ${\cal N} =8$ gauged supergravity constraints.

The constraint linking the AdS$_4$ length and the KK scale can be relaxed only in the special case where ${\widetilde m}=0$. 
In this case both the contributions from $g_6$ and the curvature are switched off and the dominant contributions become those from $g_0$ and the D6-brane sources, which must be negative to satisfy the BI constraints (see eq.~(\ref{eq:BIwithsources})).
In particular the role of giving negative energy contributions to the potential, essential for stabilization, is now played by O6-planes rather then by the curvature of the internal manifold.
The fact that such contribution scales differently with the volume and the dilaton allows to disentangle the KK scale from the AdS$_4$, indeed now
\begin{equation}
\ell_{AdS}^{-2}=\frac{V_0}{M_P^2}\sim \frac{g_0^2\sqrt{g}}{e^{-2\Phi}\sqrt{g}}\sim \frac{Q_6^2}{g_0^2 \rho^6}\,,
\end{equation}
where $Q_6$ is the net O6-plane charge contribution. In this case we have a hierarchy between the AdS$_4$ and the KK scale, which allows for a $d = 4$ effective field theory description exactly when the supergravity approximation holds, i.e.~for large volume $\rho\gg1$. 
Calabi-Yau and orbifold limit of such solution have already been discussed in \cite{vzIIA,dWGKT,Acharya:2006ne}.

Finally, notice that, unless $Q_6\gg0$, flux quantization bounds the dilaton to be such that $e^{\Phi}\lesssim 1$, forbidding the possibility of a perturbative M-theory uplift. This feature might be connected to the fact that, when the massive parameter becomes important, Type-IIA does not allow a perturbative/geometric M-theory limit anymore, so that the M-theory description is doomed to be non-geometric in this case.

\subsection{Comments on the dual CFT$_3$}

An interesting question we can ask is: what is the 3-dimensional conformal field theory (CFT) dual to this family of AdS$_4$ vacua?
We will not give the  explicit CFT but we will comment on some interesting features that can be extracted directly from the properties of the supergravity solution.

We start with the special case $m=0$, where the IIA massive deformation vanishes. 
In the absence of $g_0$, the two relevant parameters are then $g_6$ and $Q_6$, the number of D6-branes.
Notice that $Q_6$ also determines $G^{(2)}$ through the BI $dG^{(2)}=Q_6$, so that we can trade $Q_6$ with the flux of $G^{(2)}$ ($g_2$). As in \cite{ABJM}, we can be tempted to associate $g_6$ and $g_2$ with the CFT parameters $N$ and $k$, which correspond to the rank of the gauge group and the Chern-Simons (CS) level
respectively. 
Indeed, as in \cite{ABJM}, also in this case the number of colors and the 't Hooft coupling would scale with respect to the volume ($\sim \rho^6$) and the string coupling ($e^{\Phi}$) as 
\begin{equation} \label{eq:cftparam}
N\sim g_6\sim \frac{\rho^5}{e^{\Phi}}\,, \qquad \frac{N}{k}\sim \frac{g_6}{g_2}\sim \rho^4\,.
\end{equation}
If we switch on the IIA mass parameter, we expect to split the CS levels by an amount proportional to $g_0$, analogously to \cite{GT}. When ${\widetilde m}^2=15 m^2$, the net D6-brane charge vanishes and the solution becomes exact, without the need of smearing the sources. 
Notice also that in this case, as long as $\rho\gg1$, $g_0\ll g_2$, so that the splitting of the CS levels is still expected to be a small deformation of the CFT. 

In the solution without branes, the isometry group is SU(2)$^3$, which corresponds to the global flavor symmetry of the CFT.
As already noted before, adding D6/O6-brane systems corresponds to performing a ${\mathbb Z}_2$ truncation of the spectrum and to adding an SO(4)$^8$ gauge group. Analogously, the CFT is expected to be some suitable deformation of the starting CFT with global symmetries enhanced to SU(2)$^3\times$SO(4)$^8$.

A difference with respect to the CFT discussed in \cite{ABJM,GT} is the presence of 3-cycles in the supergravity solution. 
The presence of such cycles (one for each $S^3$) is associated to flat axionic directions in moduli space arising from the internal components of the RR 3-form. Consider for example
\begin{equation}\label{eq:axion}
C^{(3)}=a (\xi^1 \xi^2\xi^3+\widetilde\xi^1\widetilde\xi^2\widetilde\xi^3)\,,
\end{equation}
which is the component that survives also in the O6 case.
This field corresponds to a marginal dimension-3 operator in the gauge dual, which is a descendant of a long multiplet containing also the inverse gauge coupling field in the effective $d = 4$ supergravity.
Because of this we may expect the axion to get a mass from non-perturbative effects. 
Indeed Euclidean D2-brane instantons wrapping the two 3-spheres exactly do the job, producing corrections of the type
\begin{equation} \label{eq:E2axion}
A e^{-\int_{E2}(e^{-\Phi}{\rm Re}\,\Omega +iC^{(3)})}\sim A e^{-\frac{{\rm vol}(S^3)}{g_s}+ia} \,,
\end{equation}
where the prefactor $A$ can be in principle field-dependent.
The anomalous dimension of the dimension-3 operators associated to the axion would then get a non-perturbative correction of the type (\ref{eq:E2axion}).
If the identification of the CFT parameters (\ref{eq:cftparam}) is correct such correction would scale as
\begin{equation}
e^{-{\rm const}\sqrt{k N}}\,,
\end{equation}
thus it would be non-perturbative both in the 't Hooft coupling and in the large-$N$ expansion.

\section{Discussion}

To summarize, we studied compactifications of Type-IIA string theory on (twisted) tori with fluxes that admit a $d = 4$ description in terms of ${\cal N}=4$ supergravity. 
Since in ${\cal N}=4$ supergravity the only deformations compatible with supersymmetry are gaugings, each particular compactification will correspond to a different gauging, and each component of the possible RR, NSNS and metric fluxes that can be turned on maps into a different gauge structure constant and a different embedding into the duality group.
We thus identified the mapping between the $d = 10$ fluxes and $d = 4$ gauge structure constants. 
For the  considered class of compactifications, this allows us to reformulate the problem of finding the solutions of the $d = 10$ equation of motions to the one of finding extrema of the $d = 4$ scalar potential of the associated ${\cal N}=4$ gauged supergravity. 

This correspondence is particularly useful since there is a large number of compactifications with less supersymmetry (such as toroidal orbifolds), whose (untwisted) closed string sector is constrained 
by the underlying extended supersymmetries to be just a truncation of the ${\cal N}=4$ supergravity one.
It would be interesting to study systematically the corresponding scalar potential because it would allow us to deduce general properties valid for a large set of compactifications: for example, the (in)possibility to have full moduli stabilization in Minkowski or de Sitter space.

It is known \cite{schwei} that the gaugings of ${\cal N}=4$ supergravity include not only ``normal'' electric gaugings (associated to the structure constant $f_{+MNR}$), but also the so-called de~Roo--Wagemans phases (associated to magnetic gaugings with structure constants $f_{-MNR}$) and the Sch\"on--Weidner parameters ($\xi_{\pm M}$).
The de~Roo--Wagemans phases are essential for a complete moduli stabilization.
We identified which flux components allow us to turn on such gaugings and formulated the general rule valid also for other string compactifications. 
The Sch\"on--Weidner parameters, on the other hand, enter the scalar potential in a different way, with an intriguing similarity to Fayet--Iliopoulos terms in ${\cal N}=1$ supergravity.
We identified a $d = 10$ supergravity origin for such terms, which however does not seem compatible with a superstring uplift, for it relies on an accidental global symmetry of the two-derivative supergravity limit.
Analogously to Fayet--Iliopoulos terms in ${\cal N}=1$ supergravity, there are no known examples yet of consistent string compactifications producing non-trivial Sch\"on--Weidner parameters in four dimensions. It would be interesting to study this possibility further, because it might play an important role in the search of de~Sitter vacua in string compactifications and extended supergravities.

Another interesting direction would be the extension of our results to the inclusion of non-geometric fluxes, which would enrich the set of generated ${\cal N}=4$ gaugings.
It has recently been shown that non-geometric fluxes can produce supersymmetric Minkowski solutions with all moduli stabilized.
The extension to gaugings coming from non-geometric fluxes might in principle lead to the identification  of such vacua also in the context of ${\cal N}=4$ supergravity, a result that is still lacking in the literature.

As an application of our results, we studied the ${\cal N}=4$ uplift
of the family of supersymmetric AdS solutions found in \cite{vzIIA,cfi,AF,Behrndt:2004km,Lust:2004ig}. 
We found that for a particular choice of parameters these solutions admit a description in terms of $d=4$, ${\cal N}=4$ gauged supergravity with spontaneous supersymmetry breaking to ${\cal N}=1$.
We showed that in this case also a description in terms of ${\cal N}=8$ gauged supergravity is possible, but that there is no separation between the Kaluza--Klein and the AdS$_4$ scale, so that the gauged supergravity theory does not represent the effective $d = 4$ action, but only a consistent truncation of the $d = 10$ spectrum. 
We also showed that such solution, which corresponds to a particular $AdS_4\times S^3\times S^3$ compactification with fluxes, satisfies the $d = 10$ supersymmetry equations, which continue to be satisfied also away from the
${\cal N}=4$ point, when the solution is deformed via the introduction of sources for the D6-brane charge.
The extra parameter that control the net D6-brane charge allows to interpolate among other known IIA solutions, such as those discussed in \cite{Acharya:2003ii}.

Finally, by AdS/CFT correspondence we expect new CFT$_3$ to exist: we commented on some of their peculiar properties, which may give a hint on how to construct them.

%
%
\acknowledgments
We thank B.~Acharya, D.~Cassani, F.~Marchesano and E.~Witten for discussions. 
We also thank F.~Catino for pointing out some typos in the first version of this paper. 
This work was partially supported by the Fondazione Cariparo Excellence Grant {\em String-derived supergravities with branes and fluxes and their phenomenological implications} and by the ERC Advanced Grant no.~226455, \emph{“Supersymmetry, Quantum Gravity and Gauge Fields”} (SUPERFIELDS).
%
\appendix 
\section{Symplectic embeddings}

The $d= 4$ theory we obtained from the Scherk--Schwarz reduction of massive IIA supergravity is an ${\cal N} = 4$ gauged supergravity model.
Four-dimensional gauged supergravities are specified by their gauge group $G$ and its symplectic embedding, i.e.~the embedding of the gauge group in the electric-magnetic duality group: $G \subset {\rm Sp}(2 n_V)$, where $n_V$ is the total number of vector fields.
In this Appendix we provide the symplectic embedding specifying our model and comment on the ${\cal N}=8$ extension and on other interesting group-theoretical properties that may help to clarify the role and the origin of certain structures of the effective theory.

The starting point is the gauge group $G$ of the effective theory and its associated algebra.
For each of the vector fields $A_\mu^{\cal M} \equiv A_\mu^{\alpha M}$ we can introduce a gauge generator $T_{\cal M} \equiv T_{\alpha M}$.
These generators fulfill a gauge algebra following from the commutators 
\begin{equation}
	[T_{\cal M}, T_{\cal N}] = - X_{\cal MN}{}^{\cal P} T_{\cal P} = - X_{[{\cal MN}]}{}^{\cal P} T_{\cal P}. 
\end{equation}
We have computed in section \ref{sub:Gaugings}, eq.~(\ref{strcon}), the structure constants $f_{\alpha MN}{}^P$ of the gauge algebra realized by our model.
Following \cite{schwei}, the structure constants above are determined in terms of $f_{\alpha MNP}$ and $\xi_{\alpha M}$ as 
\begin{equation}
	\begin{array}{l}
		X_{\cal MN}{}^{\cal P} = X_{\alpha M \beta N}{}^{\gamma P} = \\[3mm]
		\displaystyle - \delta_{\beta}^\gamma f_{\alpha MN}{}^P + \frac12 \left(\delta_M^P \delta_{\beta}^\gamma \xi_{\alpha N} - \delta_N^P \delta_{\alpha}^\gamma \xi_{\beta M}- \delta_{\beta}^\gamma \eta_{MN}\xi^P_{\alpha} + \epsilon_{\alpha \beta}\delta_N^P \xi_{\delta M} \epsilon^{\delta \gamma}\right). 
	\end{array}
\end{equation}
For our model, the structure constants were derived in section \ref{sub:Gaugings} and the corresponding gauge algebra reads: 
\begin{eqnarray}
	[T_{+i}, T_{+j}] &=& \displaystyle \omega_{ij}{}^k T_{+k} - \overline H_{ij a}\, \delta^{a \bar a} \,T_{+\bar a}+ \overline G^{(6)}\, \epsilon_{ijk}\, \delta^{k\bar k}\, T_{+\bar k} - \frac12 \overline G^{(4)}_{ijab}\, \epsilon^{abc} \,T_{+c}, \label{algebrabeg}\\[2mm]
	[T_{+i}, T_{+\bar a}] &=& \displaystyle -\delta_{\bar a a}\left(\omega_{ic}{}^a T_{+c} - \overline G^{(2)}_{ib} \,\epsilon^{abc} \,T_{+c}+ \frac12\, \overline G^{(4)}_{ijbc}\, \epsilon^{abc}\, \delta^{j\bar \jmath}\, T^{+\bar \jmath}\right), \\[2mm]
	[T_{+\bar a}, T_{+b}] &=& - \omega_{ib}{}^a\, \delta_{a \bar a}\, \delta^{i \bar \imath} \, T_{+\bar \imath}, \\[2mm]
	[T_{+i}, T_{+a}] &=& \overline H_{ija}\, \delta^{i \bar \imath} \,T_{+\bar \imath}+ \omega_{ia}{}^b T_{+b}, \\[2mm]
	[T_{+i}, T_{-\bar a}] &=& \delta_{\bar a a}\left(-\frac12\, \omega_{ib}{}^a \,\delta^{b\bar b}T_{-\bar b}+\frac14 \,\epsilon_{ijk} \epsilon^{abc}\, \omega_{bc}{}^k\, \delta^{j \bar \jmath} T_{+\bar \jmath}\right), \\[2mm]
	[T_{+\bar a},T_{+\bar b}] &=& \delta_{\bar a a} \delta_{\bar b b}\left(\overline G^{(0)}\, \epsilon^{abc} T_{+c} - \overline G^{(2)}_{ic} \,\epsilon^{abc}\delta^{i\bar \imath} T_{+\bar\imath}\right), \\[2mm]
	[T_{+i}, T_{-j}] &=& - \frac14 \,\epsilon_{ijk} \,\omega_{ab}{}^k \, \epsilon^{abc} T_{+c} +\frac12 \, \omega_{ij}{}^k T_{-k} +\frac16 \, \epsilon^{abc} \, \overline H_{abc}\, \epsilon_{ijk}\, \delta^{k \bar k}\, T_{+\bar k}\nonumber \\[2mm]
	&-& \frac12 \, \overline H_{ija} \,\delta^{a \bar a} T_{-\bar a},\\[2mm]
	[T_{+i}, T_{+ \bar\imath}] &=& - \omega_{ij}{}^k\, \delta_{\bar \imath k}\, \delta^{j \bar \jmath} \,T_{+ \bar \jmath}, \\[2mm]
	[T_{+\bar a}, T_{-i}] &=& \delta_{\bar a a}\left( \frac12 \omega_{ic}{}^a\,\delta^{c \bar c}\, T_{-\bar c}- \frac14 \epsilon_{ijk} \,\omega_{bc}{}^k\, \epsilon^{abc} \,\delta^{j \bar \jmath}\, T_{+\bar \jmath}\right). \label{algebraend} 
\end{eqnarray}
This generic algebra is realized for any configuration of D6-branes and O6-planes consistent with the ${\cal N} = 4$ supersymmetry constraints.
However, when the number of D6-branes and O6-planes gives a zero net charge, the model constructed in this paper becomes a truncation of an ${\cal N} = 8$ supergravity model.
Moreover, when $\overline G^{(0)} = 0$ the model can also be obtained as an M-theory reduction with perturbative fluxes only.
For these reasons, it must be possible to embed the gauge algebra presented above into the larger ${\mathfrak e}_{7(7)}$ algebra, which is the algebra generating the U-duality group of ${\cal N} = 8$ supergravity.
We now provide this embedding explicitly.

Although the approach we use is rather indirect, it will help us clarify some interesting issues about the origin of and the constraints on the gauge group.
Our starting point is the ${\mathfrak e}_{7(7)}$ algebra.
Following \cite{deWitNicolai}, we can construct the 133 ${\mathfrak e}_{7(7)}$ generators in the fundamental \textbf{56} representation as matrices 
\begin{equation}
	T = \left(
	\begin{array}{cc}
		\delta_{[\underline{M}}^{[\underline{P}} t_{\underline{N}]}{}^{\underline{Q}]} & t_{\underline{PQRS}} \\[3mm]
		t^{\underline{MNTU}} & -\delta_{[\underline{P}}^{[\underline{T}} t_{\underline{Q}]}{}^{\underline{U}]} 
	\end{array}
	\right), 
\end{equation}
where $\underline{M},\underline{N}, \ldots = 1,\ldots,8$, $t_{\underline{M}}{}^{\underline{N}}$ are the 63 SU(8) antihermitian and traceless generators and 
\begin{equation}
	t_{\underline{MNPQ}} = \frac{1}{24} \, \epsilon_{\underline{MNPQRSTU}} \, t^{\underline{RSTU}}
\end{equation}
are the remaining 70 non-compact generators.
We then rewrite the generators and the corresponding algebra using a gl$(7,{\mathbb R})$ decomposition, which is also appropriate for M-theory embeddings.
In this basis we can split $\underline{M} = (m,8)$ and the 133 generators are $( t_{m}{}^{n}, t^{mnp}, t_{mnp}, t^{m}, t_{m} )$, as follows from the branching rule $\mathbf{133} \to \mathbf{48}_0 + \mathbf{1}_0 +\mathbf{35}_{+2}+ \overline{\mathbf{35}}_{-2} + \mathbf{7}_{-4}+ \overline{\mathbf{7}}_{+4}$.
The commutators defining the algebra then read 
\begin{eqnarray}
	\left[t_{m}{}^{n},\,t_{p}{}^{q}\right]&=&\delta_{p}^{n}\,t_{m}{}^{q}-\delta_{m}^{q}\,t_{p}{}^{n}\,,\nonumber \\[3mm]
	\left[t_{m}{}^{n},\,t^{p_1p_2p_3}\right]&=&\displaystyle-3\,\delta_{m}^{[p_1}\,t^{p_2p_3]n}+\frac{5}{7}\,\delta_{m}^{n}\,t^{p_1p_2p_3}\,,\nonumber \\[3mm]
	\left[t_{m}{}^{n},\,t_{p}\right]&=&\displaystyle\delta_{p}^{n}\,t_{m}+\frac{3}{7}\,\delta_{m}^{n}\,t_{p}\,,\nonumber \\[3mm]
	\left[t^{n_1n_2n_3},\,t^{p_1p_2p_3}\right]&=&\epsilon^{n_1n_2n_3p_1p_2p_3q }\,t_{q}\,,\nonumber \\[3mm]
	\left[t_{m}{}^{n},\,t_{p_1p_2p_3}\right]&=&\displaystyle 3\,\delta_{[p_1}^{n}\,t_{p_2p_3]m}-\frac{5}{7}\,\delta_{m}^{n}\,t_{p_1p_2p_3}\,,\nonumber \\[3mm]
	\left[t_{m}{}^{n},\,t^{p}\right]&=&\displaystyle-\delta_{m}^{p}\,t^{n}-\frac{3}{7}\,\delta_{m}^{n}\,t^{p}\,, \\[3mm]
	\left[t_{n_1n_2n_3},\,t_{p_1p_2p_3}\right]&=&\epsilon_{n_1n_2n_3p_1p_2p_3q}\,t^{q}\,,\nonumber \\[3mm]
	\left[t^{n},\,t_{m}\right]&=&\displaystyle t_{m}{}^{n}+\frac{1}{7}\,\delta_{m}^{n}\,t\,,\nonumber \\[3mm]
	\left[t^{m},\,t^{n_1n_2n_3}\right]&=&\displaystyle -\frac{1}{6}\,\epsilon^{mn_1n_2n_3p_1p_2p_3}\,t_{p_1p_2p_3}\,,\nonumber \\[3mm]
	\left[t_{m},\,t_{n_1n_2n_3}\right]&=&\displaystyle -\frac{1}{6}\,\epsilon_{mn_1n_2n_3p_1p_2p_3}\,t^{p_1p_2p_3}\,,\nonumber \\[3mm]
	\left[t_{m_1m_2m_3},\,t^{n_1n_2n_3}\right]&=&\displaystyle 18\,\delta^{[n_1n_2}_{[m_1m_2}\,t_{m_3]}{}^{n_3]}-\frac{24}{7}\,\delta^{n_1n_2n_3}_{m_1m_2m_3}\,t\,, \nonumber 
\end{eqnarray}
where $t\equiv t_{m}{}^{m}$.
We realized this splitting because whenever $\overline G^{(0)} = 0$ the gauge algebra (\ref{algebrabeg})--(\ref{algebraend}) reduces to the one obtained from M-theory compactifications with geometric fluxes, 4-form fluxes $G_{mnpq}$ and a 7-form flux $G^{(7)}$, and although this uplift can be done only when the IIA mass parameter is switched off, the ${\cal N} = 8$ embedding can still be performed in the presence of non-trivial $\overline G^{(0)}$.

In the M-theory framework, the 56 vector fields and their corresponding generators also split as $\mathbf{56} \to \overline\mathbf{7}_{-3} + \mathbf{21}_{-1} + \overline\mathbf{21}_{+1} + \mathbf{7}_{+3}$.
We can actually label them as the ones coming from the reduction of the metric fields ($V_\mu^m$) $Z_m$, the ones associated to the 3-form fields ($A^{(3)}_{\mu mn}$) $W^{mn}$, the dual ones coming from the 6-form ($A^{(6)}_{\mu pqrst}$) $W_{mn}$ and the dual metric generators ($\widetilde V_{\mu m}$) $Z^m$.
These generators can now be embedded in the ${\mathfrak e}_{7(7)}$ ones by recognizing the fluxes as intertwiners between the representations of the generators and those of the ${\mathfrak e}_{7(7)}$ generators.
The identification of the M-theory perturbative fluxes in terms of our IIA fluxes is straightforward.
The 4-form, the geometric fluxes and the 6-form flux proportional to the volume of the internal space lift to objects of the same type (where the volume of the internal space is now 7-dimensional): 
\begin{equation}
	\overline G^{(4)}_{ijab}\,, \quad \omega_{ij}{}^k, \quad \omega_{ia}{}^b, \quad \omega_{ab}{}^k, \quad G^{(7)} = \overline G^{(6)}\,. 
\end{equation}
The other fields can also be identified easily as 
\begin{equation}
	\omega_{ia}{}^{11} = \overline G^{(2)}_{ia}, \quad G_{11\,ija} = \overline H_{ija}\,, \quad G_{11\,abc} = \overline H_{abc}\,. 
\end{equation}
We are left with a single non-perturbative flux $\overline G^{(0)}$, which, however, can also be easily identified by looking at the structure of the commutators of the gauge algebra as a component of a flux in the $\mathbf{28}_{+1}$ (see for instance section 4 of \cite{DallAgata:2007sr}): 
\begin{equation}
	\xi^{mn} = \overline G^{(0)} \, \delta^m_7 \delta^n_7. 
\end{equation}
At this stage we can propose the embedding of the M-theory generators in the ${\mathfrak e}_{7(7)}$ ones as 
\begin{eqnarray}
	Z_m &=& a_1 \, \omega_{mn}{}^p\, t_p{}^n + a_2\, G_{mnpq}\, t^{npq} + a_3\, g_6\, t_m, \label{Zmd}\\[2mm]
	W^{mn} &=& 2 b_1 \, \omega_{pq}{}^{[m}t^{n]pq} + b_2 \, \epsilon^{mnpqrsv} G_{pqrs} t_v + 2 b_3 \, \xi^{p[m} t_p{}^{n]}, \label{Wmnu}\\[2mm]
	W_{mn} &=& c_1 \omega_{mn}{}^p \, t_p, \label{Wmnd}\\[2mm]
	Z^m &=& d_1 \xi^{mn}\, t_n, \label{Zmu} 
\end{eqnarray}
leading to the embedding tensors 
\begin{equation}
	\begin{array}{rclcrclcrcl}
		\theta_{m,n}{}^p &=& a_1 \omega_{mn}{}^p, &\phantom{sp}& \theta_{m,npq} &=& a_2 G_{mnpq}, &\phantom{sp}& \theta_{m,}{}^n &=& a_3 g_6 \delta_m^n, \\[2mm]
		\theta_{mn,}{}^p &=& c_1 \omega_{mn}{}^p, \\[2mm]
		\theta^{mn,}{}_{p}{}^q &=& 2 b_3 \xi^{q[m}\delta_p^{n]}, &\phantom{sp}& \theta^{mn,}{}_{pqr} &=& 2 b_1 \omega_{[pq}{}^{[m}\delta^{n]}_{r]}, &\phantom{sp}& \theta^{mn,p} &=& b_2 \epsilon^{mnpqrsu} G_{qrsu}, \\[2mm]
		\theta^{m,n} &=& d_1 \xi^{mn}. 
	\end{array}
\end{equation}

For the gauging to be well defined, these tensors must satisfy some quadratic constraints: 
\begin{eqnarray}
	\theta_{m,}{}^p \theta^{m,q} + \theta^{mn,p} \theta_{mn,}{}^q - \theta^{m,p}\theta_{m,}{}^q + \theta_{mn,}{}^p \theta^{mn,q} = 0, && \label{quad1}\\[2mm]
	\theta^{m,p} \theta_{m,qrs} + \theta_{mn,}{}^p \theta^{mn,}{}_{qrs} = 0, && \label{quad2}\\[2mm]
	\theta^{m,p} \theta_{m,q}{}^r + \theta_{mn,}{}^p \theta^{mn,}{}_q{}^r = 0.\label{quad3} 
\end{eqnarray}
It is straightforward to show that (\ref{quad3}) is identically satisfied, while (\ref{quad1}) corresponds to the 3-form BI, and (\ref{quad2}) gives the 4-form BI and the torsion constraints $\omega \cdot \omega = 0$.

Hence we can finally derive the structure of the gauge algebra defined by the generators (\ref{Zmd})--(\ref{Zmu}): 
\begin{eqnarray}
	[Z_m,Z_n] &=& \omega_{mn}{}^p Z_p + \beta \, G_{mnpq} W^{pq} + \gamma \, g_6 W_{mn}, \\[2mm]
	[Z_m,W^{np}] &=& 2 \delta \, \omega_{mq}{}^{[n} W^{p]q} + \varepsilon\, \epsilon^{npq_1 q_2 q_3 q_4 q_5} G_{m q_1 q_2 q_3} W_{q_4 q_5} + 2 \chi \,g_6 \delta_m^{[n} Z^{p]},\\[2mm]
	[Z_m,W_{np}] &=& \zeta\, \omega_{np}{}^q W_{mq}, \\[2mm]
	[W^{mn},W^{pq}] &=& -4 \eta\, \xi^{[m[p} W^{q]n]} + 2 \phi\, \epsilon^{pq r_1 r_2 r_3 r_4 [m} G_{r_1 r_2 r_3 r_4} Z^{n]}, \\[2mm]
	[W^{mn},W_{pq}] &=& 2 \sigma\, \omega_{pq}{}^{[m} Z^{n]}, 
\end{eqnarray}
with all the other commutators vanishing identically.
Closure in ${\mathfrak e}_{7(7)}$ through the definitions (\ref{Zmd})--(\ref{Zmu}) fixes the various coefficients to 
\begin{equation}
	\begin{array}{l}
		\displaystyle \beta = \frac32 \frac{a_2}{b_1},\quad \gamma = \frac{a_3}{c_1}, \quad \delta = 1, \quad \varepsilon = - \frac{a_2 b_1}{c_1}, \\
		\displaystyle \chi = \frac{a_3}{2 c_1}, \quad \zeta = 1, \quad \rho = - \frac{b_1^2}{2 c_1}, \quad \eta = - \frac23 \frac{b_1}{a_2}, \\[2mm]
		\displaystyle \phi = - \frac{a_2 b_1}{4 c_1}, \quad \sigma = - \frac12, 
	\end{array}
\end{equation}
and
\begin{equation}
	b_3 = \frac{b_1}{3 a_2}, \quad b_2 = \frac{a_2 b_1}{2}, \quad d_1 = \frac23 \, \frac{c_1 b_1}{a_2}. 
\end{equation}

Obviously we cannot have 56 independent generators and a simple inspection of (\ref{Zmd})--(\ref{Zmu}) immediately confirms this, leading to the following constraints: 
\begin{equation}
	3 \omega_{[mn}{}^q W_{|q|p]} = \frac32 \frac{a_2}{b_1} G_{mnpq} Z^q, 
\end{equation}
\begin{equation}
	\omega_{pq}{}^m W^{pq} = \frac{a_2 b_1}{2 c_1} \epsilon^{m n_1n_2n_3n_4n_5n_6} G_{n_1n_2n_3n_4}W_{n_5n_6} + \frac{a_3}{c_1} Z^m - \frac23 \frac{b_1}{a_2} \xi^{mn} Z_n. 
\end{equation}

At this stage, following \cite{dagfer}, we can deduce how the action of the ${\cal N} = 4$ gauge generators can be embedded in ${\mathfrak e}_{7(7)}$ in the case without net D6-brane charge, according to the branching of the representations of ${\mathfrak e}_{7(7)}$ with respect to o$(1,1)^3 \times {\rm sl}(3) \times {\rm sl}(3)$.
In particular, from the branching of the $\mathbf{56}$ we get that the surviving 24 vectors transform as 
\begin{eqnarray}
	&&(\bar{\mathbf{3}}, \mathbf{1})_{---} +(\bar{\mathbf{3}}, \mathbf{1})_{+--} + (\mathbf{1},{\mathbf{3}})_{-0-}+(\mathbf{1},\bar{\mathbf{3}})_{-0+} \nonumber \\ &+&
	({\mathbf{3}}, \mathbf{1})_{+++} +({\mathbf{3}}, \mathbf{1})_{-++}+ (\mathbf{1},\bar{\mathbf{3}})_{+0+}+(\mathbf{1},{\mathbf{3}})_{+0-}\,, 
\end{eqnarray}
which is the representation content of our vector fields 
\begin{equation}
	V_\mu^i, \quad C_{\mu ij}, \quad B_{\mu a}, \quad C_{\mu ab}, \quad \widetilde V_{\mu i}, \quad C_{\mu abc i}, \quad \widetilde B_{\mu ijk ab}, \quad C_{\mu ijk a}  \,,
\end{equation}
and of the corresponding generators 
\begin{equation}
	T_{+i}, \quad T_{-i}, \quad T_{+\bar a}, \quad T_{+a}, \quad T_{-\bar \imath}, \qquad T_{+\bar \imath}, \qquad T_{-a}, \qquad T_{-\bar a}. \  {}
\end{equation}
We can then proceed to embed the gauge generators in the ones of ${\mathfrak e}_{7(7)}$ using the fluxes as intertwineres and splitting the indices as $m=(i,a,11)$.
The result is 
\begin{eqnarray}
	T_{+i} &\equiv& \omega_{ij}{}^k t_k{}^j+\omega_{ia}{}^b t_b{}^a + \overline G^{(2)}_{ia} t_{11}{}^a +\frac12 \overline G^{(4)}_{ijab}t^{jab} -\overline H_{ija}t^{ja 11} - \overline G^{(6)} \, t_i,\\[3mm]
	T_{+a} &\equiv& -\epsilon_{abc}\omega_{id}{}^{b} t^{cid} +\frac12 \epsilon^{ijk} \overline H_{aij} t_k, \\[2mm]
	T_{+\bar \imath} &\equiv& - \delta_{\bar \imath i} \frac12 \epsilon^{ijk}\omega_{jk}{}^l t_l, \\[2mm]
	T_{+\bar a} &\equiv&\delta_{\bar a a}\left( \omega_{ic}{}^a t^{11 ic} - \overline G^{(2)}_{ic} t^{aic} +\frac14 \epsilon^{ijk} \epsilon^{abc} \overline G^{(4)}_{ijbc} t_k - \overline G^{(0)} t_{11}{}^a\right), \\[2mm]
	T_{-i} &\equiv& -\frac12 \epsilon_{ijk} \omega_{ab}{}^{j} t^{kab} + \frac{1}{6} \epsilon^{abc} \overline H_{abc} t_i, \\[2mm]
	T_{-\bar a} &\equiv& \frac12 \delta_{\bar a a} \epsilon^{abc}\omega_{bc}{}^k t_k. 
\end{eqnarray}

As we have seen before, not all gauge vectors will be independent, therefore the corresponding gauge generators will be constrained.
For the case at hand, in the absence of net D6-brane charge, the constraints follow from the above embedding in ${\mathfrak e}_{7(7)}$: 
\begin{equation}
	- \omega_{ab}{}^k \epsilon_{ijk} \delta^{j \bar \jmath} T_{+ \bar \jmath} + \omega_{i[a}{}^c \epsilon_{b]cd} \, \delta^{d \bar d} \, T_{-\bar d}= 0, \label{constra1} 
\end{equation}
\begin{equation}
	\omega_{ij}{}^k \,\epsilon^{ijl}\, T_{-l} + \omega_{ab}{}^k\, \epsilon^{abc}\, T_{+c} + \frac{1}{3}\, \epsilon^{abc}\, \overline H_{abc} \,\delta^{k \bar k} \,T_{+\bar k} - \epsilon^{ijk} \,\overline H_{ija} \, \delta^{a \bar a}\, T_{-\bar a}= 0. \label{constra2} 
\end{equation}
This fact has an interesting application in the context of understanding the process by which we have identified the electric vector fields and integrated out the magnetic ones.
Indeed, the above constraints are in one-to-one correspondence with the linear combinations of the BI that have to be solved to obtain the physical vector fields, without introducing two-form tensor fields in the $d= 4$ effective theory.
For this purpose we can take as a starting point the massive IIA action where both the standard and the dual field-strengths appear.
We then solve the BI resulting from the integration of the potentials we do not want in the effective action.
These BI read 
\begin{equation}
	d({\rm e}^B G) = 0. \label{generalBianchi} 
\end{equation}
The standard formulation of the effective theory can be obtained by integrating out $C^{(9)}$, $C^{(7)}$ and $C^{(5)}$, but by doing so, we get an effective ${\cal N}=4$ supergravity model with tensor fields: $C_{\mu\nu\rho}$ and $C_{\mu\nu i}$.
If we do not want tensor fields in the effective $d= 4$ theory, we have to integrate out $C^{(9)}$, $C^{(7)}$, and some components of $C^{(5)}$ \emph{together with} some components of $C^{(3)}$.
This means that we have to solve the BI for the 4-form and 6-form only partially.
We therefore need to identify which combinations of the BI have to be selected.
This can be done in the following way.
Start by taking the BI coming from integrating out the $C^{(p-1)}$ potentials and define 
\begin{equation}
	d G^{(p)} + \omega G^{(p)} + H \,G^{(p-2)} \equiv F^{(p+1)}, 
\end{equation}
where $H = d B + \omega B + \overline H$. Trivial consistency conditions are 
\begin{equation}
	d F^{(p+1)} + H \, F^{(p-1)} + B \, d F^{(p-1)} = 0. \label{consistency} 
\end{equation}
The parameterizations of the curvatures are obtained by first integrating out $C^{(9)}$ and $C^{(7)}$, leading to $F^{(1)} = 0$ and $F^{(3)} = 0$.
This results in the definition of the $\overline G^{(0)}$ flux and of the curvature two-form $G^{(2)} = d C^{(1)} + \omega C^{(1)} + \overline G^{(2)} - B \overline G^{(0)}$.
However, when we proceed to the integration of the 5-form, we solve the Bianchi identities corresponding only to some of the components of $C^{(5)}$.
These are $C_{\mu\nu\rho\sigma a}$, $C_{\mu\nu\rho ia}$, $C_{\mu\nu abc}$ and $C_{\mu\nu a ij}$, which correspond to all the forms of rank greater than one.
These should not appear in the effective theory.
On the other hand we do not want to integrate out the scalar fields $C_{abcij}$ and we have to decide which components of the vector fields $C_{\mu abci}$ and $C_{\mu ijk a}$ have to survive.
Their minimal set is now easily determined by imposing the consistency conditions (\ref{consistency}).
If we want to solve $F_{\mu ij ab} = 0$, $F_{\mu\nu ijk} = 0$ and $F_{\mu\nu iab} = 0$ (corresponding to the 5-form tensor fields with rank $>1$), we also need to solve at least some of the Bianchi identities related to the 5-form vector fields because of the consistency conditions 
\begin{equation}
	(d F^{(5)})_{\mu\nu\rho ijk} = 0, \qquad (d F^{(5)})_{\mu\nu\rho iab} = 0. 
\end{equation}
Upon using $F_{\mu\nu ijk} = 0$ and $F_{\mu\nu iab} = 0$, these consistency conditions read 
\begin{equation}
	3\, \omega_{[ij}{}^l F_{\mu\nu\rho k] l} = 0, \label{ibF1} 
\end{equation}
which is identically vanishing when $\omega_{ij}{}^j = 0$, and 
\begin{equation}
	\omega_{ab}{}^l F_{\mu\nu\rho i l} + 2\,\omega_{i[a}{}^c F_{\mu\nu\rho b]c} = 0. \label{ibF2} 
\end{equation}
These equations are selecting the linear combinations related to the tensor fields we have integrated out.
Moreover they are in one-to-one correspondence with the constraints (\ref{constra1}) on the corresponding gauge generators.
It is easy to check that the combinations appearing in (\ref{ibF2}) do not contain any tensor fields and hence we can solve $G_{\mu\nu ij}$ and $G_{\mu\nu ab}$ in terms of vector fields only.

At this point we can move to the integration of the 3-form degrees of freedom we do not want to see in the effective action.
This means the space-time 3-form $C_{\mu\nu\rho}$, the three 2-forms $C_{\mu\nu i}$ and consequently the (up to 3) vector fields selected by the same mechanism as the one described above.
The integration of the 3 tensor fields $C_{\mu\nu i}$ implies that $F_{\mu\nu ijabc} = 0$.
However, the consistency condition now reads 
\begin{equation}
	d F^{(7)} + \omega F^{(7)} + H \, F^{(5)} = 0, 
\end{equation}
because we did not solve all the equations from $F^{(5)} =0$, but only some of them.
Looking at the 3 directions labeled by $\mu\nu\rho ij abc$ we get that 
\begin{equation}
	\begin{array}{l}
		\omega_{ij}{}^l F_{\mu\nu\rho l abc} + 3 \omega_{[ab}{}^l F_{\mu\nu\rho |l ij|c]} + 6 \omega_{[i|[a}{}^d F_{\mu\nu\rho bc]j]d} + \\[2mm]
		+ (\overline H_{abc} + 3 \omega_{[ab}{}^l B_{c]l}) F_{\mu\nu\rho ij} + 3 (\overline H_{ij[a} - \omega_{ij}{}^l B_{l[a]} + 2\omega_{a][i]}{}^c B_{j]c}) F_{\mu\nu\rho bc]} = 0. 
	\end{array}
\end{equation}
We can see once more that only some parts of the vector field Bianchi identities participate in the above conditions and once more they are in one-to-one correspondence with the constraints (\ref{constra2}).
  


\end{document}